\documentclass[twocolumn,preprintnumbers,amsmath,amssymb,prl,
reprint,groupedaddress,amsmath,citeautoscript,flushbottom,
floatfix,superscriptaddress]{revtex4}
\usepackage{graphicx}% Include figure files
\usepackage{dcolumn}% Align table columns on decimal point
\usepackage{bm}% bold math
\usepackage{amsmath}
\usepackage{amssymb}
\usepackage{amsmath}
\usepackage{booktabs}
\usepackage{multirow}
\usepackage{array}
\usepackage{color}
\usepackage{xcolor}
\usepackage[normalem]{ulem}
\usepackage{float}
\makeatletter
\def\@hangfrom@section#1#2#3{\@hangfrom{#1#2#3}}
\makeatother
\newcommand{\vc}[1]{\boldsymbol{#1}}

\begin{document}

\title{Kitaev Spin Liquid in 3$\boldsymbol{d}$ Transition Metal Compounds}
\author{Huimei Liu}
\affiliation{Max Planck Institute for Solid State Research,
Heisenbergstrasse 1, D-70569 Stuttgart, Germany}

\author{Ji\v{r}\'{\i} Chaloupka}
\affiliation{Department of Condensed Matter Physics, Faculty of Science,
Masaryk University, Kotl\'a\v{r}sk\'a 2, 61137 Brno, Czech Republic}
\affiliation{Central European Institute of Technology, Masaryk University,
Kamenice 753/5, 62500 Brno, Czech Republic}

\author{Giniyat Khaliullin}
\affiliation{Max Planck Institute for Solid State Research,
Heisenbergstrasse 1, D-70569 Stuttgart, Germany}

\begin{abstract}
We study the exchange interactions and resulting magnetic phases in the honeycomb cobaltates. For a broad range of trigonal crystal fields acting on Co$^{2+}$ ions, the low-energy pseudospin-1/2 Hamiltonian is dominated by bond-dependent Ising couplings that constitute the Kitaev model. The non-Kitaev terms nearly vanish at small values of trigonal field $\Delta$, resulting in spin liquid ground state. Considering Na$_3$Co$_2$SbO$_6$ as an example, we find that this compound is proximate to a Kitaev spin liquid phase, and can be driven into it by slightly reducing $\Delta$ by $\sim 20$~meV, e.g., via strain or pressure control. We argue that due to the more localized nature of the magnetic electrons in 3$d$ compounds, cobaltates offer the most promising search area for Kitaev model physics.

\end{abstract}

\date{\today}
\maketitle
The Kitaev honeycomb model~\cite{Kit06}, demonstrating the key concepts of quantum spin liquids~\cite{Sav17} via an elegant exact solution, has attracted much attention (see the recent reviews~\cite{Her18,Tre17,Win17_m,Tak19,Mot20}). In this model, the nearest-neighbor (NN) spins $S=1/2$ interact via a simple Ising-type coupling $S_i^\gamma S_j^\gamma$. However, the Ising axis $\gamma$ is not global but bond-dependent, taking the mutually orthogonal directions ($x,y,z$) on the three adjacent NN-bonds on the honeycomb lattice. Having no unique easy-axis and being frustrated, the Ising spins fail to order and form instead a highly entangled quantum many-body state, supporting fractional excitations described by Majorana fermions~\cite{Kit06}.

Much effort has been made to realize the Kitaev spin liquid (SL) experimentally. From a materials perspective, the Ising-type anisotropy is a hallmark of unquenched orbital magnetism. As the orbitals are spatially anisotropic and bond-directional, they naturally lead to the desired bond-dependent exchange anisotropy via spin-orbit coupling~\cite{Kha05}. Along these lines, 5$d$ iridates have been suggested~\cite{Jac09} to host Kitaev model; later, 4$d$ RuCl$_3$ was added~\cite{Plu14} to the list of candidates. To date, however, the Kitaev SL remains elusive, as this state is fragile and destroyed by various perturbations, such as small admixture of a conventional Heisenberg coupling~\cite{Cha10} caused by direct overlap of the $d$ orbitals. Even more detrimental to Kitaev SL are the longer range couplings~\cite{Win16_m}, unavoidable in weakly localized 5$d$- and 4$d$-electron systems with the spatially extended $d$ wave functions. We thus turn to 3$d$ systems with more compact $d$ orbitals~\cite{note_r2}.

While the idea of extending the search area to 3$d$ materials is appealing, and plausible theoretically~\cite{Liu18_m,San18}, it raises an immediate question crucial for experiment: Is spin-orbit coupling (SOC) in 3$d$ ions strong enough to support the orbital magnetism prerequisite for the Kitaev model design? This is a serious concern, since noncubic crystal fields present in real materials tend to quench orbital moments and suppress the bond-dependence of the exchange couplings~\cite{Kha05}. In this Letter, we give a positive answer to this question. Our quantitative analysis of the crystal field effects on the magnetism of 3$d$ cobaltates shows that the orbital moments remain active and generate a Kitaev model as the leading term in the Hamiltonian. In fact, we identify the trigonal crystal field as the key and experimentally tunable parameter, which decides the strength of the non-Kitaev terms in 3$d$ compounds.

%--------------------------------------------------------------------------
\begin{figure} [b]
\begin{center}
\includegraphics[width=8.3cm]{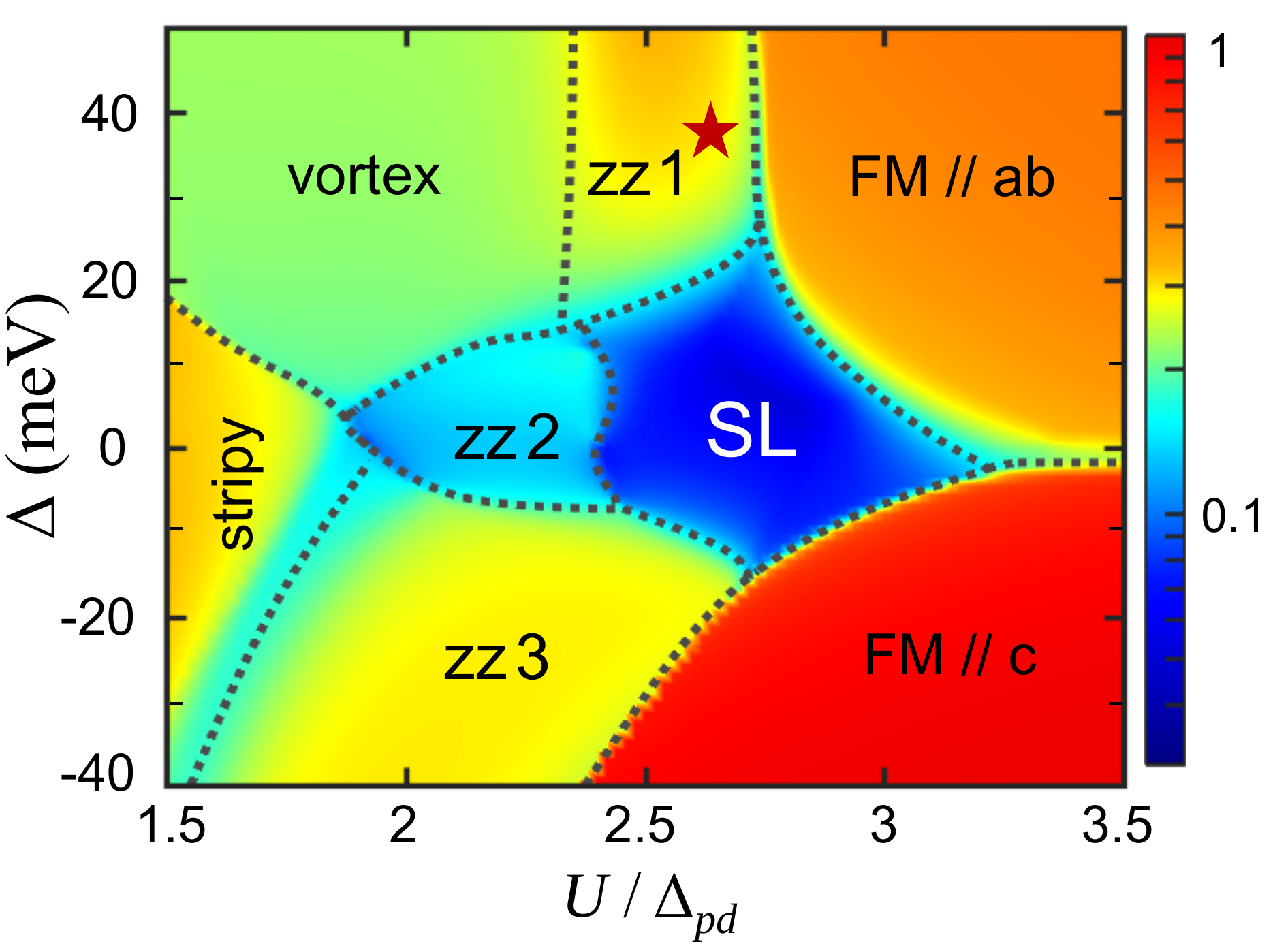}
\caption{The calculated magnetic phase diagram of honeycomb cobaltates. The Kitaev SL phase is surrounded by ferromagnetic (FM) states with moments in the honeycomb $ab$-plane and along the $c$-axis, zigzag-type states with moments in the $ab$ plane (zz1), along Co-O bonds (zz2), and in the $ac$ plane (zz3). Vortex- and stripy-type phases take over at smaller $U/\Delta_{pd}$. The color map shows the second-NN spin correlation strength (leading eigenvalue of the correlation matrix $\langle \widetilde{S}^\alpha_i \widetilde{S}^\beta_j\rangle$ normalized by $\widetilde{S}^2=1/4$), which drops sharply in the SL phase. The star indicates the rough position of Na$_3$Co$_2$SbO$_6$.
}
\label{fig:1}
\end{center}
\end{figure}
%--------------------------------------------------------------------------

Our main results are summarized in Fig.~\ref{fig:1}, displaying various magnetic phases of spin-orbit entangled pseudospin-1/2 Co$^{2+}$ ions on a honeycomb lattice. The phase diagram is shown as a function of trigonal field $\Delta$, in a window relevant for honeycomb cobaltates, and a ratio of Coulomb repulsion $U$ and the charge-transfer gap $\Delta_{pd}$~\cite{Zaa85}. From the analysis of experimental data, we find that Na$_3$Co$_2$SbO$_6$~\cite{Vic07_m,Won16_m,Yan19_m} is located at just $\sim 20$~meV ``distance'' from the Kitaev SL phase (see Fig.~1), and could be driven there by a $c$-axis compression that reduces $\Delta$. This seems feasible, given that $\Delta$ variations within a window of $\sim 70$~meV were achieved by strain control in a cobalt oxide~\cite{Csi05}.

We now describe our calculations resulting in Fig.~\ref{fig:1}. In short, we first derive the pseudospin exchange interactions from a microscopic theory, as a function of various parameters, and then obtain the corresponding ground states numerically by exact diagonalization.

{\it Exchange interactions}.-- In an octahedral environment, Co$^{2+}$ ion with $t_{2g}^5e_g^2$ configuration possesses spin $S=3/2$ and effective orbital moment $L=1$, which form, via spin-orbit coupling, a pseudospin $\widetilde{S}=1/2$~\cite{Abr70_m}. Over decades, cobaltates served as a paradigm for quantum magnetism, providing a variety of pseudospin-1/2 models ranging from the Heisenberg model in perovskites with corner-sharing octahedra~\cite{Hol71,Buy71} to the Ising model when the CoO$_6$ octahedra share their edges~\cite{Col10}.

\begin{figure}
\begin{center}
\includegraphics[width=8.5cm]{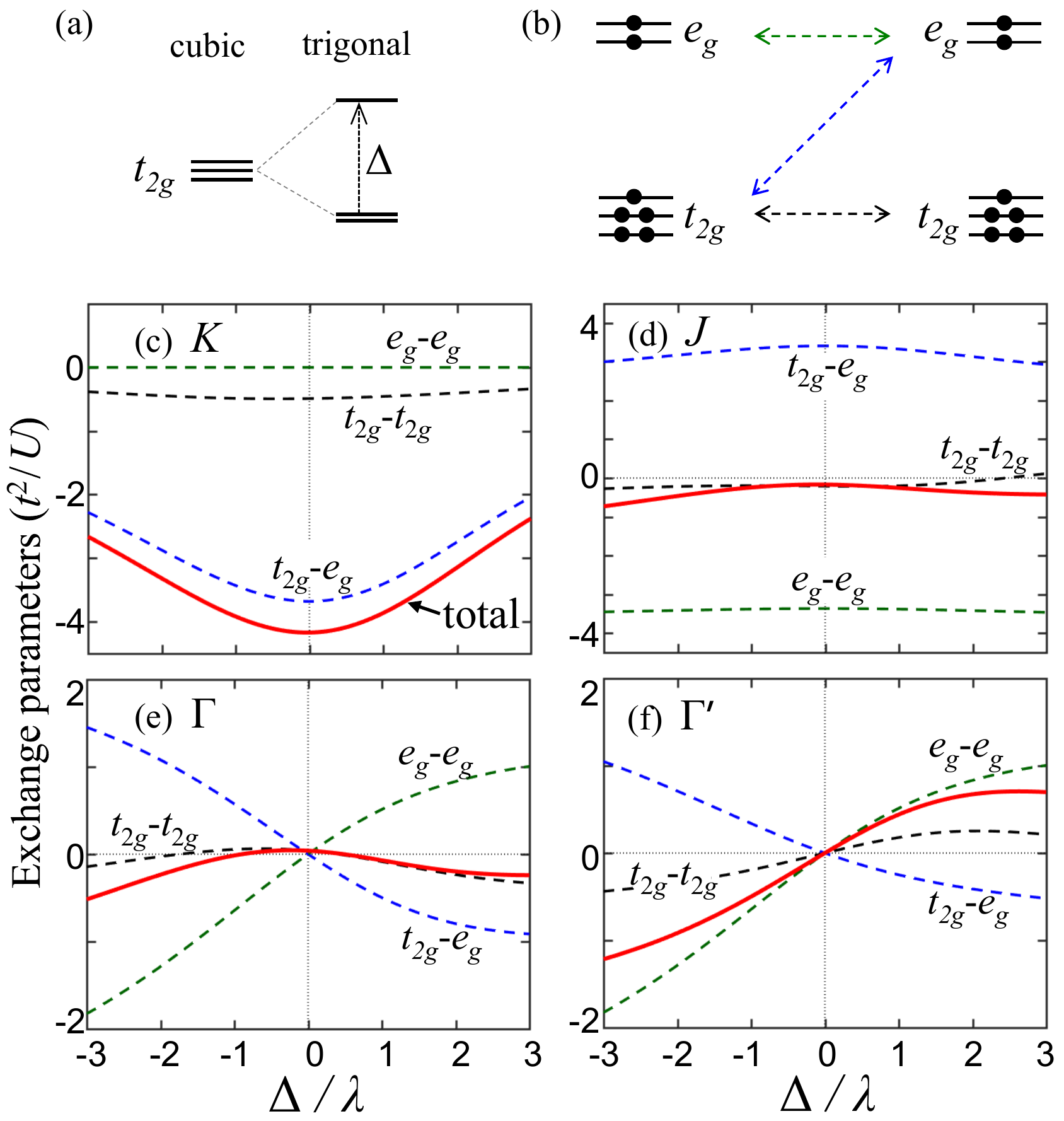}
\caption{(a) Splitting of $t_{2g}$-electron level under trigonal crystal field. (b) Schematic of the spin-orbital exchange channels for $d^7$ ions. (c)-(f) Exchange parameters $K$, $J$, $\Gamma$, and $\Gamma'$ (red solid lines) as a function of $\Delta/\lambda$, calculated at $U/\Delta_{pd}=2.5$ and Hund's coupling $J_H=0.15U$. On each panel, dashed lines show individual contributions of $t_{2g}$-$t_{2g}$ (black), $t_{2g}$-$e_g$ (blue), and $e_g$-$e_g$ (green) exchange channels. The couplings $J$, $\Gamma$, and $\Gamma'$ nearly vanish in the cubic limit $\Delta=0$.
}
\label{fig:2}
\end{center}
\end{figure}

A microscopic theory of Co$^{2+}$ interactions in the edge-sharing geometry has been developed just recently~\cite{Liu18_m,San18}, assuming an ideal cubic symmetry. Here we consider a realistic case of trigonally distorted lattices, where $t_{2g}$ orbitals split as shown in Fig.~\ref{fig:2}(a). Our goal is to see if such distortions leave enough room for the Kitaev model physics in real compounds. This is decided by the spin-orbital structure of the pseudospin $\widetilde{S}=1/2$ wave functions; in terms of $|S_Z,L_Z \rangle$ states (the trigonal axis $Z \!\parallel \!c$ is perpendicular to the honeycomb plane), they read as:
\begin{equation}
\Big|\! \pm \widetilde{\frac{1}{2}}\Big\rangle =
\mathcal{C}_1\Big| \pm \frac{3}{2}, \mp 1\Big\rangle+
\mathcal{C}_2\Big| \pm \frac{1}{2},0\Big\rangle+
\mathcal{C}_3\Big| \mp \frac{1}{2},\pm 1\Big\rangle.
\label{eq:wf_m}
\end{equation}
The coefficients $\mathcal{C}_{1,2,3}$ depend on a relative strength $\Delta/\lambda$ of the trigonal field $\Delta(L^2_Z-\tfrac{2}{3})$ and SOC $\lambda \vc L \cdot \vc S$~\cite{Lin63_m,SM}. At $\Delta=0$, one has $(\mathcal{C}_1,\mathcal{C}_2,\mathcal{C}_3) =(\tfrac{1}{\sqrt{2}},\tfrac{-1}{\sqrt{3}},\tfrac{1}{\sqrt{6}})$, and all the three components of $\vc L$ are equally active. A positive (negative) $\Delta$ field tends to quench $L_Z$ ($L_{X/Y}$).

The next step is to project various spin-orbital exchange interactions in cobaltates~\cite{Liu18_m} onto the above pseudospin-1/2 subspace. The calculations are standard but very lengthy; the readers interested in details are referred to the Supplemental Material~\cite{SM}. At the end, we obtain the $\widetilde{S}=1/2$ Kitaev model $K \widetilde{S}_i^\gamma \widetilde{S}_j^\gamma$, supplemented by Heisenberg $J$ and off-diagonal anisotropy $\Gamma, \Gamma'$ terms; for $\gamma=z$ type NN bonds, they read as:
\begin{align}
\mathcal{H}_{ij}^{(z)}=K \widetilde{S}_i^z\widetilde{S}_j^z +
J  \widetilde{\vc S}_i \cdot  \widetilde{\vc S}_j +
\Gamma (\widetilde{S}_i^x\widetilde{S}_j^y +\widetilde{S}_i^y\widetilde{S}_j^x &)
\notag \\
+\; \Gamma'(\widetilde{S}_i^x\widetilde{S}_j^z +
\widetilde{S}_i^z\widetilde{S}_j^x+
\widetilde{S}_i^y\widetilde{S}_j^z +
\widetilde{S}_i^z\widetilde{S}_j^y &).
\label{eq:H_m}
\end{align}
Interactions $\mathcal{H}_{ij}^{(\gamma)}$ for $\gamma=x,y$ type bonds follow from a cyclic permutation among $\widetilde{S}_j^x$, $\widetilde{S}_j^y$, and $\widetilde{S}_j^z$.

While the Hamiltonian (\ref{eq:H_m}) is of the same form as in $d^5$ Ir/Ru systems~\cite{Win17_m,Cha15_m}, the microscopic origin of its parameters $K, J, \Gamma, \Gamma'$ is completely different in $d^7$ Co compounds. This is due to the spin-active $e_g$ electrons of Co$(t_{2g}^5e_g^2)$ ions, which generate new spin-orbital exchange channels $t_{2g}$-$e_g$ and $e_g$-$e_g$, shown in Fig.~\ref{fig:2}(b), in addition to the $t_{2g}$-$t_{2g}$ ones operating in $d^5$ systems with $t_{2g}$-only electrons. In fact, the new terms make a major contribution to the exchange parameters, as illustrated in Figs.~\ref{fig:2}(c)-\ref{fig:2}(f). In particular, Kitaev coupling $K$ comes almost entirely from the $t_{2g}$-$e_g$ process. It is also noticed that $t_{2g}$-$e_g$ and $e_g$-$e_g$ contributions to $J$, $\Gamma$, and $\Gamma'$ are of opposite signs and largely cancel each other, resulting in only small overall values of these couplings.

%--------------------------------------------------------------------------
%--------------------------------------------------------------------------

Figure~\ref{fig:2} shows that the trigonal field $\Delta$, which acts via modification of the pseudospin wavefunction (\ref{eq:wf_m}), has an especially strong impact on the non-Kitaev couplings $J$, $\Gamma$, $\Gamma'$. As a result, the relative strength ($J/K$, etc) of these ``undesired'' terms is very sensitive to $\Delta$ variations. This suggests the orbital splitting $\Delta$ as an efficient (and experimentally accessible) parameter that controls the proximity of cobaltates to the Kitaev-model regime.

\begin{figure}
\begin{center}
\includegraphics[width=8.5cm]{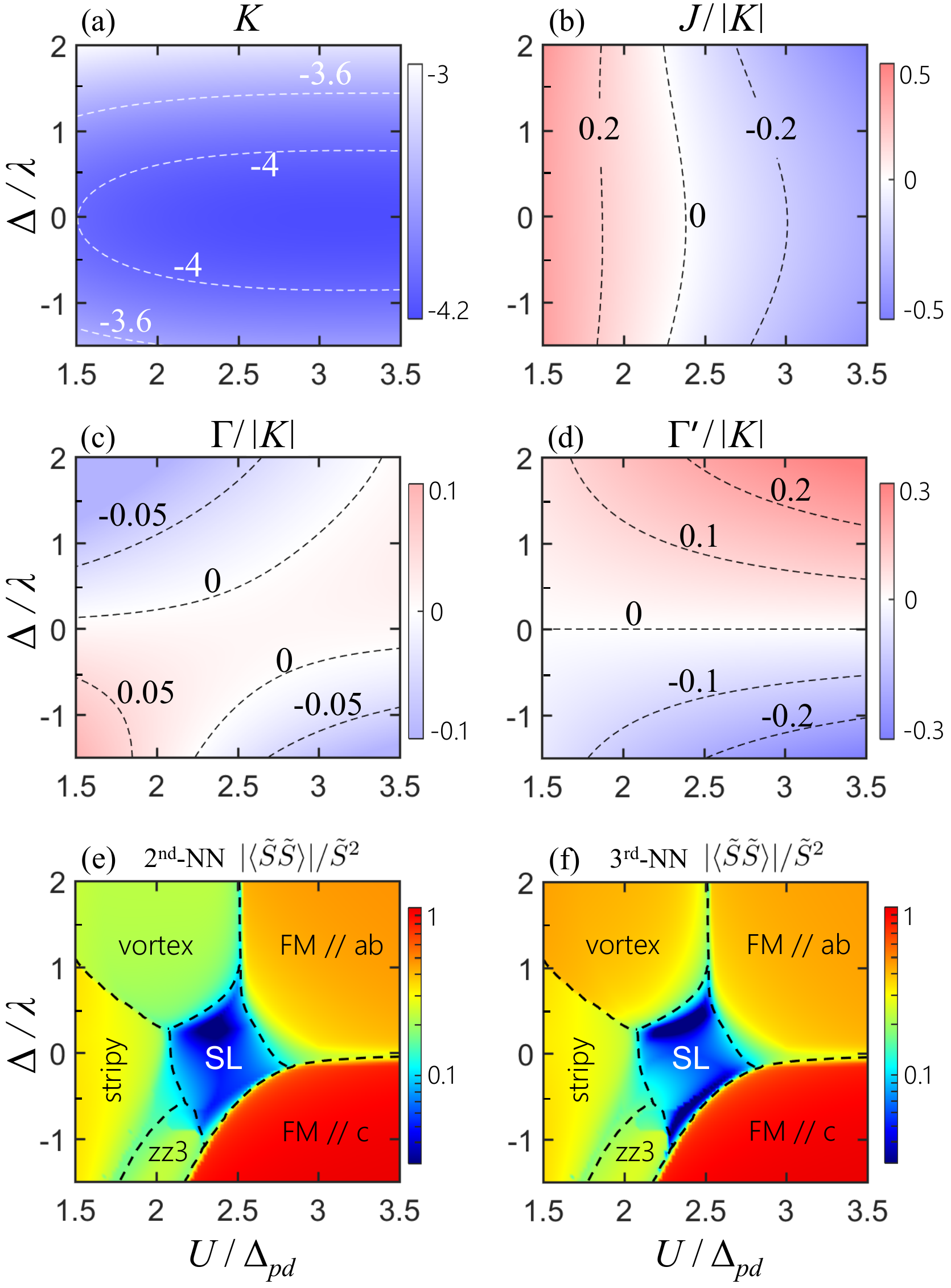}
\caption{(a) Kitaev coupling $K$ (in units of $t^2/U$), and (b)-(d) the relative values of $J/|K|$, $\Gamma/|K|$, and $\Gamma'/|K|$ as a function of $\Delta/\lambda$ and $U/\Delta_{pd}$. For convenience, specific values of parameters are indicated by contour lines. (e)-(f) The corresponding phase diagram obtained by ED of the model on a hexagon-shaped 24-site cluster. As in Fig.~\ref{fig:1}, the color maps quantify the strength of (e) second-NN and (f) third-NN spin correlations, which drop sharply in the SL phase (small but finite values are due to deviations from the pure Kitaev model~\cite{Cha10}).
}
\label{fig:3}
\end{center}
\end{figure}

Another important parameter in the theory is the $U/\Delta_{pd}$ ratio. In contrast to Ir/Ru-based Mott insulators with small $U/\Delta_{pd}\sim 0.5$, cobaltates are charge-transfer insulators~\cite{Zaa85}, with typical values of $U/\Delta_{pd}\sim 2-3$ depending on the material chemistry. Including both Mott-Hubbard $U$ and charge-transfer $\Delta_{pd}$ excitations, we have calculated~\cite{SM} the exchange couplings as a function of $U/\Delta_{pd}$ and $\Delta/\lambda$. Figure \ref{fig:3}(a) shows that Kitaev coupling $K$ is not much sensitive to $U/\Delta_{pd}$ variations. On the other hand, the non-Kitaev terms, especially Heisenberg coupling $J$, are quite sensitive to $U/\Delta_{pd}$, see Figs.~\ref{fig:3}(b)-\ref{fig:3}(d). However, their values relative to $K$ remain small over a broad range of parameters.

{\it Phase diagram}.-- Having quantified the exchange parameters in Hamiltonian (\ref{eq:H_m}), we are now ready to address the corresponding ground states. As Kitaev coupling is the leading term, the model is highly frustrated. We therefore employ exact diagonalization (ED) which has been widely used to study phase behavior of the extended Kitaev-Heisenberg models (see, e.g., Refs.~\cite{Cha10,Cha13_m,Oka13,Rau14,Cha16_m,Rus19_m}). In particular, by utilizing the method of coherent spin states~\cite{Cha16_m,Rus19_m}, we can detect and identify the magnetically ordered phases (including easy-axis directions for the ordered moments). When non-Kitaev couplings are small (roughly below $10\%$ of the FM $K$ value), a quantum spin-liquid state is expected. Reflecting the unique feature of the Kitaev model~\cite{Kit06}, this state is characterized by short-range spin correlations that are vanishingly small beyond nearest-neighbors~\cite{Cha10}.

The resulting phase diagram, along with the data quantifying spin correlations beyond NN distances, is presented in Figs.~\ref{fig:3}(e) and \ref{fig:3}(f). The main trends in the phase map are easy to understand considering the variations of non-Kitaev couplings with $\Delta/\lambda$ and $U/\Delta_{pd}$. As we see in Figs.~\ref{fig:3}(c) and \ref{fig:3}(d), $\Gamma'$ exactly vanishes at the $\Delta=0$ line, and $\Gamma$ is very small too. Thus, in the cubic limit, the model (\ref{eq:H_m}) essentially becomes the well studied $K-J$ model, with large FM Kitaev $K$ term, and $J$ correction changing from AF $J>0$ to FM $J<0$ as a function of $U/\Delta_{pd}$. Consequently, the ground state changes from stripy AF (at small $U/\Delta_{pd}$) to FM order at large $U/\Delta_{pd}$, through the Kitaev SL phase in between~\cite{Cha13_m}. In the SL phase, spin correlations are indeed short-ranged and bond-selective: for $z$-type NN bonds, we find $\langle \widetilde{S}^z \widetilde{S}^z \rangle/\widetilde{S}^2\simeq 0.52$ (as in the Kitaev model), while they nearly vanish at farther distances, see Figs.~\ref{fig:3}(e) and \ref{fig:3}(f).

As we switch on the trigonal field $\Delta$, the $\Gamma'$ term comes into play confining the SL phase to the window of $|\Delta|/\lambda<1$ (where $|\Gamma'/K|<0.1$). In the FM phases, the sign of $\Gamma'$ decides the direction of the FM moments. On the left-top (left-bottom) part of the phase map, where Heisenberg coupling $J$ is AF, the stripy state gives way to a vortex-type~\cite{Cha15_m} (zigzag-type) ordering, stabilized by the combined effect of $\Gamma$ and $\Gamma'$ terms.

%--------------------------------------------------------------------------
%--------------------------------------------------------------------------

To summarize up to now, the nearest-neighbor pseudospin Hamiltonian is dominated by the FM Kitaev model, which appears to be robust against trigonal splitting of orbitals. Subleading terms, represented mostly by $J$ and $\Gamma'$ couplings, shape the phase diagram, which includes a sizeable SL area. While these observations are encouraging, it is crucial to inspect how the picture is modified by longer range interactions, especially by the third-NN Heisenberg coupling $J_3\widetilde{\vc S}_i\cdot \widetilde{\vc S}_j$, which appears to be one of the major obstacles on the way to a Kitaev SL in 5$d$ and 4$d$ compounds~\cite{Win16_m,Win17_m}. We have no reliable estimate for $J_3$, since long-range interactions involve multiple exchange channels and are thus sensitive to material chemistry details. As such, they have to be determined experimentally. We note that $|J_3/K|\simeq 0.1$ was estimated~\cite{Win18,Win17a} in the 4$d$ compound RuCl$_3$; in cobaltates with more localized 3$d$ orbitals~\cite{note_r2}, this ratio is expected to be smaller.

Adding a $J_3$ term to the model (\ref{eq:H_m}), we have re-examined the ground states and found that the Kitaev SL phase is stable up to $|J_3/K|\sim 0.06$~\cite{SM}. The modified phase diagram, obtained for a representative value of $J_3=0.15 t^2/U \simeq 0.04|K|$, is shown in Fig.~\ref{fig:1}~\cite{note_axis}. Its comparison with Fig.~\ref{fig:3} tells that the main effect of $J_3$ is to support the zigzag-type states (with different orientation of moments) at the expense of other phases. Note also that the SL area is shifted to the right, where FM $J$ and AF $J_3$ tend to frustrate each other. The phase diagram in Fig.~\ref{fig:1} should be generic to Co$^{2+}$ honeycomb systems, and will be used in the following discussion.

{\it Honeycomb lattice cobaltates}.-- A number of such compounds are known: $A_3$Co$_2$SbO$_6$ ($A$=Na,Ag,Li)~\cite{Vic07_m,Won16_m,Yan19_m,Zve16_m,Str19_m}, Na$_2$Co$_2$TeO$_6$~\cite{Vic07_m,Lef16_m,Ber17_m,Yao19_m}, BaCo$_2$($X$O$_4$)$_2$ ($X$=As, P)~\cite{Reg06_m,Zho19_m,Nai18_m,Zho18_m}, CoTiO$_3$~\cite{New64_m,Bal17_m,Yua19_m}, CoPS$_3$~\cite{Bre86_m,Wil17_m}. They are quasi-two-dimensional magnets; within the $ab$-planes, zigzag or FM order is most common.

Traditionally, experimental data in Co$^{2+}$ compounds is analysed in terms of an effective $\widetilde{S}=1/2$ models of $XXZ$ type~\cite{Hut73_m,Reg06_m,Tom11_m,Ros17_m,Nai18_m,Yua19_m}. As $\widetilde{S}=1/2$ magnons ($\sim 10$~meV) are well separated from higher lying spin-orbit excitations ($\sim 30$~meV), the pseudospin picture itself is well justified; however, a conventional $XXZ$ model neglects the bond-directional nature of pseudospin $\widetilde{S}=1/2$ interactions. A general message of our work is that a proper description of magnetism in cobaltates should be based on the model of Eq.~(\ref{eq:H_m}), supplemented by longer-range interactions. We note in passing that the $XXZ$ model also follows from Eq.~(\ref{eq:H_m}) when the Kitaev-type anisotropy is suppressed~\cite{Cha15_m}; however, such an extreme limit is unlikely for realistic trigonal fields, given the robustness of the $K$ coupling, see Fig.~\ref{fig:3}.

As an example, we consider Na$_3$Co$_2$SbO$_6$ which has low N\'{e}el temperature and a reduced ordered moment~\cite{Yan19_m}. Analysing the magnetic susceptibility data~\cite{Yan19_m} including all spin-orbit levels~\cite{SM}, we obtain a positive trigonal field $\Delta \simeq 38$~meV and $\lambda \simeq 28$~meV; these values are typical for Co$^{2+}$ ions in an octahedral environment (see, e.g., Ref.~\cite{Yua19_m}). With $\Delta/\lambda \simeq 1.36$, we evaluate $\widetilde{S}=1/2$ doublet $g$-factors $g_{ab}\simeq 4.6$ and $g_c\simeq 3$, from which a saturated moment of $2.3\mu_B$, consistent with the magnetization data~\cite{Yan19_m}, follows.

Zigzag-ordered moments in Na$_3$Co$_2$SbO$_6$ are confined to the $ab$ plane~\cite{Yan19_m}; this corresponds to the zz1 phase in Fig.~\ref{fig:1}. The easy-plane anisotropy is due to the $\Gamma'$ term, which is positive for $\Delta>0$, see Fig.~\ref{fig:3}(d). Regarding the location of Na$_3$Co$_2$SbO$_6$ on the $U/\Delta_{pd}$ axis of Fig.~\ref{fig:1}, we believe it is close to the FM/\!/$ab$ phase, based on the following observations. First, a sister compound Li$_3$Co$_2$SbO$_6$ has $ab$-plane FM order~\cite{Str19_m} (most likely due to smaller Co-O-Co bond angle, $91^{\circ}$ versus $93^{\circ}$, slightly enhancing the FM $J$ value). Second, zigzag order gives way to fully polarized state at small magnetic fields~\cite{Vic07_m,Yan19_m}. These facts imply that zz1 and FM/\!/$ab$ states are closely competing in Na$_3$Co$_2$SbO$_6$.

\begin{figure}
\begin{center}
\includegraphics[width=8.5cm]{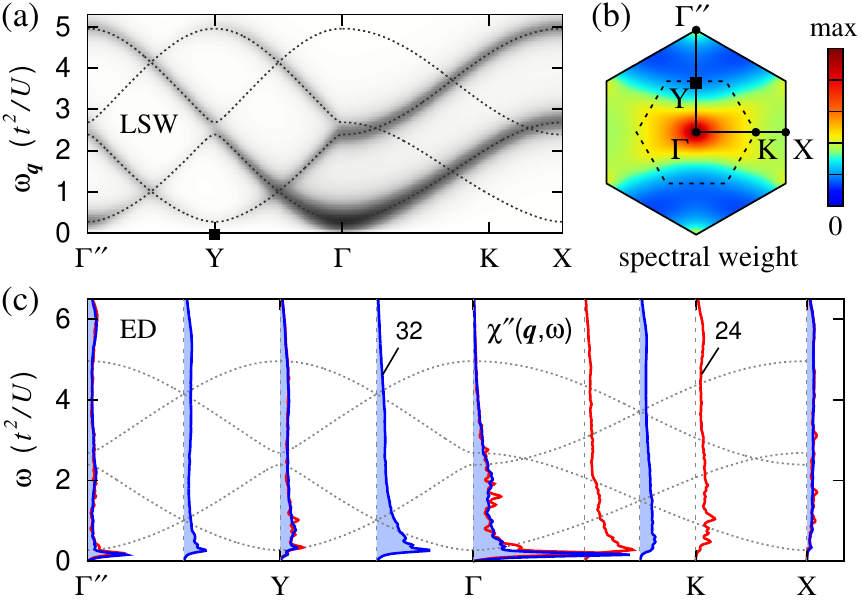}
\caption{Spin excitation spectrum expected in Na$_3$Co$_2$SbO$_6$. The parameters $K=-3.6$, $J=-0.5$, $\Gamma=-0.1$, $\Gamma'=0.6$ (in units of $t^2/U$) follow from our theory, while $J_3=0.15$ is added ``by hand''~\cite{note_J3} to stabilize the zigzag order. (a) Magnon dispersions and intensities from linear spin wave (LSW) theory. (b) The energy-integrated magnon intensity over the Brillouin zone. The intensity is largest around $\Gamma$, i.e. away from the Bragg point Y. (c) Exact diagonalization results for hexagonal 24- and 32-site clusters. Plotted is the trace $\chi''(\vc q,\omega)$ of the spin susceptibility tensor~\cite{SM}, which comprises the low-energy magnon peak and a broad continuum.
}
\label{fig:4}
\end{center}
\end{figure}

Based on the above considerations, we roughly locate Na$_3$Co$_2$SbO$_6$ in the phase diagram as shown in Fig.~\ref{fig:1}. In this parameter area, the exchange couplings are $K\simeq -3.6\;t^2/U$, $J/|K|\sim -0.14$, $\Gamma/|K|\sim -0.03$, and $\Gamma'/|K|\sim 0.16$, see Figs.~\ref{fig:3}(a)-\ref{fig:3}(d). The small values of $J,\Gamma,\Gamma'$ imply the proximity to the Kitaev model, explaining a strong reduction of the ordered moments from their saturated values~\cite{Yan19_m}. As a crucial test for our theory, we show in Fig.~\ref{fig:4} the expected spin excitations. The large FM Kitaev interaction enhances magnon spectral weight near $\vc q=0$ and leads to its anisotropy in momentum space, see Figs.~\ref{fig:4}(a) and \ref{fig:4}(b). The ED results in Fig.~\ref{fig:4}(c) show that, as a consequence of the dominant Kitaev coupling, magnons are strongly renormalized and only survive at low energies, and a broad continuum of excitations~\cite{Win17a,Goh17} as in RuCl$_3$~\cite{Ban18,San15} emerges. Neutron scattering experiments on Na$_3$Co$_2$SbO$_6$ are desired to verify these predictions.

%--------------------------------------------------------------------------
%--------------------------------------------------------------------------

If the above picture is confirmed by experiments, the next step should be to drive Na$_3$Co$_2$SbO$_6$ into the Kitaev SL state. As suggested by Fig.~\ref{fig:1}, this requires a reduction of the trigonal field by $\sim 20$~meV, e.g. by means of strain or pressure control. At this point, the relative smallness of SOC for 3$d$ Co ions comes as a great advantage: while strong enough to form the pseudospin  moments, it makes the lattice manipulation of the $\widetilde{S}=1/2$ wave functions (and hence magnetism) far easier than in iridates~\cite{Liu19}. Monitoring the magnetic behavior of Na$_3$Co$_2$SbO$_6$ and other honeycomb cobaltates under uniaxial pressure would be thus very interesting.

To conclude, we have presented a comprehensive theory of exchange interactions in honeycomb cobaltates, and studied their magnetic phase behavior. The analysis of Na$_3$Co$_2$SbO$_6$ data suggests that this compound is proximate to a Kitaev SL phase and could be driven there by a $c$-axis compression. A broader message is that as one goes from 5$d$ Ir to 4$d$ Ru and further to 3$d$ Co, magnetic $d$ orbitals become more localized, and this should improve the conditions for realization of the nearest-neighbor-only interaction model designed by Kitaev.

We thank A. Yaresko, T. Takayama, and A. Smerald for discussions, and M. Songvilay for sharing unpublished data. G.Kh. acknowledges support by the European Research Council under Advanced Grant No. 669550 (Com4Com). J.Ch. acknowledges support by Czech Science Foundation (GA\v{C}R) under Project No.~GA19-16937S and M\v{S}MT \v{C}R under NPU II project CEITEC 2020 (LQ1601). Computational resources were supplied by the project ``e-Infrastruktura CZ'' (e-INFRA LM2018140) provided within the program Projects of Large Research, Development and Innovations Infrastructures.

% ==================================================

\clearpage
%\vspace{10cm}
\onecolumngrid

\setcounter{equation}{0}
\setcounter{figure}{0}
\setcounter{table}{0}
\setcounter{page}{7}
\makeatletter
\renewcommand{\thesection}{S\Roman{section}}
\renewcommand{\thetable}{S\arabic{table}}
\renewcommand{\theequation}{S\arabic{equation}}
\renewcommand{\thefigure}{S\arabic{figure}}

\begin{center}
\textbf{\large Supplemental Material for\\
Kitaev Spin Liquid in $3d$ Transition Metal Compounds}
\end{center}
\setcounter{equation}{0}
\setcounter{figure}{0}
\setcounter{table}{0}
\setcounter{page}{7}
\makeatletter
\renewcommand{\thesection}{S\Roman{section}}
\renewcommand{\thetable}{S\arabic{table}}
\renewcommand{\theequation}{S\arabic{equation}}
\renewcommand{\thefigure}{S\arabic{figure}}

%=======================================================================
\section{I. Single-ion wavefunctions}
%=======================================================================

\begin{figure}
\begin{center}
\includegraphics[width=12cm]{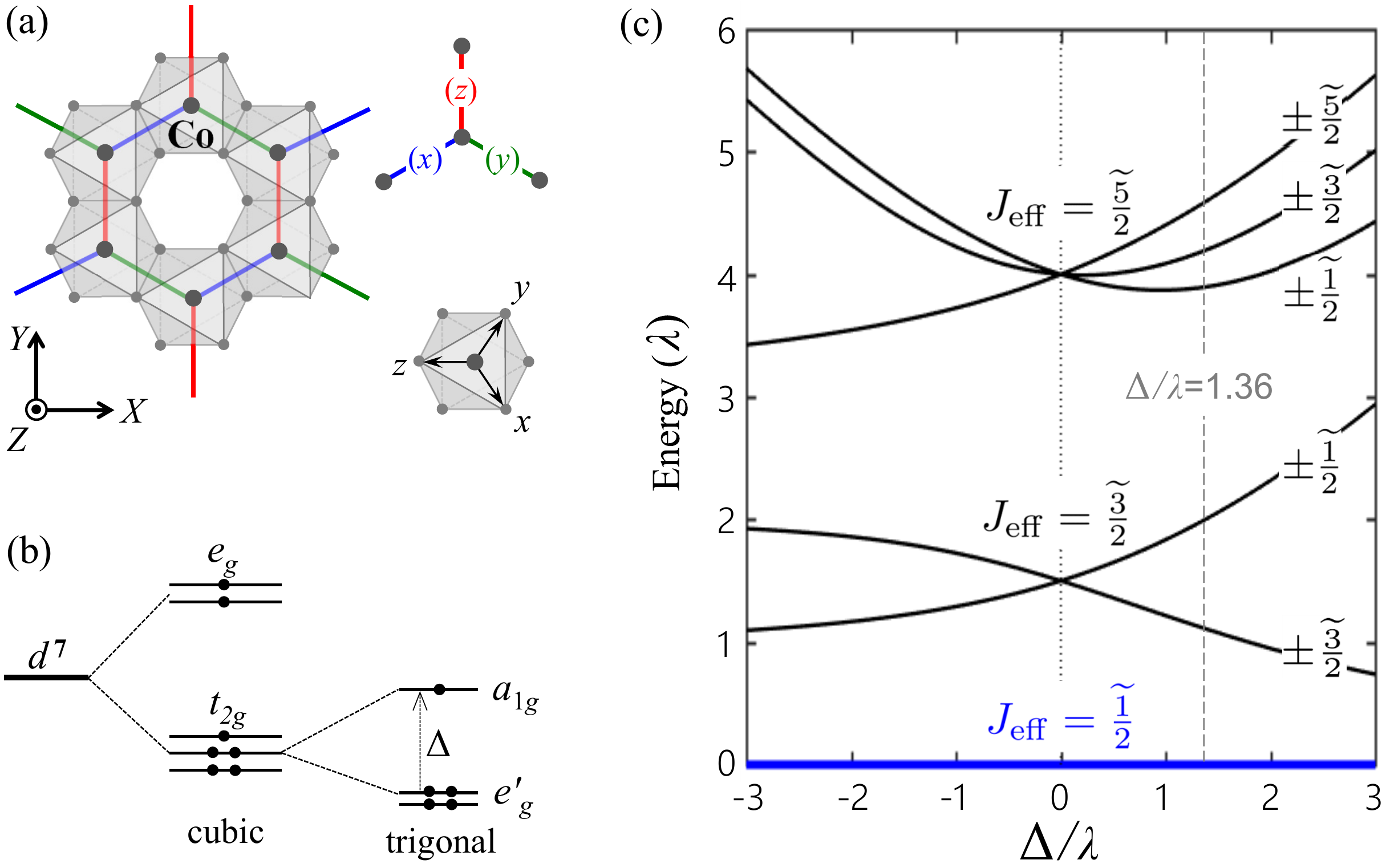}
\caption{(a) Top view of the honeycomb cobaltate plane, $x$, $y$, and $z$ type NN-bonds are shown in blue, green, and red colors, respectively. The definition of global $X$, $Y$, $Z$ and the local cubic $x$, $y$, $z$ axes are shown in insets. (b) High-spin $d^7(t_{2g}^5e_g^2)$ configuration in the trigonal crystal field $\Delta$. (c) Splitting of $S=3/2$, $L=1$ manifold of Co$^{2+}$ ion under spin-orbit coupling $\lambda$ and trigonal field $\Delta$. At $\Delta/\lambda=1.36$ (appropriate for Na$_3$Co$_2$SbO$_6$), the first excited state energy is about $\lambda\sim 30$ meV.}
\label{fig:s1}
\end{center}
\end{figure}

The $d^7$ Co$^{2+}$ ions in an octahedral crystal field have predominantly $t_{2g}^5e_g^2$ configuration with a high spin $S=3/2$~\cite{Abr70}. A trigonal distortion along $Z$-axis splits the $t_{2g}$ manifold into an orbital singlet $a_{1g}$ and a doublet $e_g'$ by energy $\Delta$, see Fig.~\ref{fig:s1}(a,b). In the electron representation, it is captured by the Hamiltonian
$H_{\Delta}=\tfrac{1}{3}\Delta (2n_{a_{1g}}-n_{e'_g})$.
In terms of the effective angular momentum $L=1$ of the Co$^{2+}$ ions,
the $a_{1g}$-hole configuration corresponds to $L_Z=0$, while the $e_g'$
doublet hosts the $L_Z=\pm 1$ states. Consequently, the trigonal field Hamiltonian translates into $H_{\Delta}=\Delta (L_Z^2-\tfrac{2}{3})$.
The following relations between the $L$-states and orbitals hold:
\begin{align}
|L_Z=0\rangle &=\frac{1}{\sqrt{3}} \left(|a\rangle+|b\rangle+|c\rangle \right),
\notag\\
|L_Z= \pm 1\rangle &= \pm \frac{1}{\sqrt{3}} \left(e^{ \pm i\tfrac{2\pi}{3}}|a\rangle
+e^{\mp i\tfrac{2\pi}{3}}|b\rangle+|c\rangle \right),
\end{align}
where shorthand notations $a=d_{yz}$, $b=d_{zx}$, and $c=d_{xy}$ are used.

Diagonalization of $H_{\Delta}=\Delta (L_Z^2-\tfrac{2}{3})$ and $H_{\lambda}=\lambda \vc L \cdot \vc S$ results in a level structure shown in Fig.~\ref{fig:s1}(c). The states are labeled according to the total angular momentum $J_{\rm eff}=\widetilde{\frac{1}{2}}$, $\widetilde{\frac{3}{2}}$, and $\widetilde{\frac{5}{2}}$.
The ground state Kramers doublet hosts a pseudospin $\widetilde{S}=1/2$; its wavefunctions, written in the basis of $|S_Z,L_Z \rangle$, read as:
\begin{align}
\Big| \!\widetilde{\frac{1}{2}},\pm \widetilde{\frac{1}{2}} \Big\rangle =
\mathcal{C}_1 \Big| \pm\frac{3}{2}, \mp 1 \Big\rangle + \mathcal{C}_2 \Big|
\pm \frac{1}{2},0 \Big\rangle+\mathcal{C}_3\Big| \mp \frac{1}{2}, \pm 1 \Big\rangle .
\label{eq:wf}
\end{align}
The coefficients obey a relation $\mathcal{C}_1:\mathcal{C}_2:\mathcal{C}_3 =
\frac{\sqrt{6}}{r_1}:-1:\frac{\sqrt{8}}{r_1+2}$, where the parameter $r_1>0$  is determined by the equation $\frac{\Delta}{\lambda}
=\frac{r_1+3}{2}-\frac{3}{r_1}-\frac{4}{r_1+2}$ \cite{Lin63}.
The ground state energy is
\begin{align}
E_{\rm GS}=\frac{\Delta}{3}-\frac{\lambda}{2}(r_1+3).
\label{eq:en}
\end{align}
The exchange Hamiltonian between the pseudospins $\widetilde{S}=1/2$ is obtained by projecting the Kugel-Khomskii type spin-orbital Hamiltonians onto the ground state doublet (\ref{eq:wf}).

We also specify the excited states, which will be needed in Sec. IV to calculate the magnetic susceptibility. The wavefunctions and energies for
$\Big|\!\widetilde{\frac{3}{2}},\pm \widetilde{\frac{1}{2}}\Big\rangle$
and
$\Big|\!\widetilde{\frac{5}{2}}, \pm \widetilde{\frac{1}{2}}\Big\rangle$
states share the same form as of Eq.~\ref{eq:wf} and Eq.~\ref{eq:en}, but with
different $r_1$. Namely, the above equation $\frac{\Delta}{\lambda} =\frac{r_1+3}{2}-\frac{3}{r_1}-\frac{4}{r_1+2}$ has three roots. The root $r_1>0$ corresponds to the ground state. The other two roots with $-2<r_1<0$ and $r_1<-2$ correspond to
$\Big|\!\widetilde{\frac{3}{2}},\pm \widetilde{\frac{1}{2}}\Big\rangle$
and
$\Big|\!\widetilde{\frac{5}{2}}, \pm \widetilde{\frac{1}{2}}\Big\rangle$ states, respectively. The wavefunctions and energies of the remaining states are:
\begin{alignat}{2}
&\Big| \!\widetilde{\frac{3}{2}},\pm \widetilde{\frac{3}{2}} \Big\rangle =
 c_{\varphi}\Big| \pm \frac{3}{2},0 \Big\rangle
- s_{\varphi}\Big| \pm \frac{1}{2}, \pm 1 \Big\rangle \;,
&\quad \ \ \ \ \ \ \ \ \ \ \ \ \
&E( \!\widetilde{\tfrac{3}{2}},\pm \widetilde{\tfrac{3}{2}} )
=-\tfrac{1}{2} \sqrt{\left( \Delta+ \tfrac{1}{2} \lambda \right)^2+6\lambda^2}
+ \tfrac{1}{4} \lambda -  \tfrac{1}{6} \Delta  \; ,
\notag \\
&\Big| \!\widetilde{\frac{5}{2}},\pm \widetilde{\frac{3}{2}} \Big\rangle =
s_{\varphi}\Big| \pm  \frac{3}{2},0 \Big\rangle
+c_{\varphi}\Big| \pm  \frac{1}{2}, \pm 1 \Big\rangle \;,
 &
&E( \!\widetilde{\tfrac{5}{2}},\pm \widetilde{\tfrac{3}{2}})
=\tfrac{1}{2} \sqrt{\left( \Delta+ \tfrac{1}{2} \lambda \right)^2+6\lambda^2}
+ \tfrac{1}{4} \lambda -  \tfrac{1}{6} \Delta  \; ,
\notag \\
&\Big| \!\widetilde{\frac{5}{2}},\pm \widetilde{\frac{5}{2}} \Big\rangle =
\Big| \pm \frac{3}{2},\pm 1\Big\rangle \;, &
&E( \!\widetilde{\tfrac{5}{2}},\pm \widetilde{\tfrac{5}{2}})
=\tfrac{3}{2} \lambda +\tfrac{2}{3} \Delta  \; .
\label{eq:Tr}
\end{alignat}
Here, $c_{\varphi}=\cos \varphi$, $s_{\varphi}=\sin \varphi$, and
$\tan 2\varphi =2\sqrt{6}\lambda/(2\Delta+\lambda)$.

%--------------------------------------------------------------------------
%--------------------------------------------------------------------------

%=======================================================================
\section{II. Pseudospin $\widetilde{S}=1/2$ Hamiltonian and calculation of its parameters}
%=======================================================================

\label{sec:eH}
In the cubic reference frame, pseudospin-1/2 interactions on $z$-type bonds have a general form
\begin{align}
\mathcal{H}_{ij}^{(z)}= K \widetilde{S}_i^z\widetilde{S}_j^z +
J  \widetilde{\vc S}_i \cdot  \widetilde{\vc S}_j
+ \Gamma (\widetilde{S}_i^x\widetilde{S}_j^y +
\widetilde{S}_i^y\widetilde{S}_j^x )
+ \Gamma'(\widetilde{S}_i^x\widetilde{S}_j^z +
\widetilde{S}_i^z\widetilde{S}_j^x+
\widetilde{S}_i^y\widetilde{S}_j^z +
\widetilde{S}_i^z\widetilde{S}_j^y) \;.
\label{eq:HK}
\end{align}
The interactions on $x$ and $y$ type bonds are obtained by cyclic permutations among $\widetilde{S}_j^x$, $\widetilde{S}_j^y$, and $\widetilde{S}_j^z$.

The Hamiltonian in Eq.~\ref{eq:HK} can also be written in global $XYZ$ reference
frame \cite{Cha15}:
\begin{align}
\mathcal{H}_{ij}^{(\gamma)}= & J_{XY}
\left(\widetilde{S}_i^X\widetilde{S}_j^X +
\widetilde{S}_i^Y\widetilde{S}_j^Y \right) +
J_Z \widetilde{S}_i^Z\widetilde{S}_j^Z  \notag \\
+ & A \left[ c_{\gamma}\left(\widetilde{S}_i^X\widetilde{S}_j^X -
\widetilde{S}_i^Y\widetilde{S}_j^Y \right)-
s_{\gamma}\left(\widetilde{S}_i^X\widetilde{S}_j^Y +
\widetilde{S}_i^Y\widetilde{S}_j^X \right) \right]   \notag \\
- & B \sqrt {2}\left[ c_{\gamma}
\left(\widetilde{S}_i^X\widetilde{S}_j^Z +
\widetilde{S}_i^Z\widetilde{S}_j^X \right)
+s_{\gamma}\left(\widetilde{S}_i^Y\widetilde{S}_j^Z +
\widetilde{S}_i^Z\widetilde{S}_j^Y \right) \right] ,
\label{eq:H}
\end{align}
with $c_{\gamma}\equiv \cos \phi_{\gamma}$ and $s_{\gamma}\equiv \sin \phi_{\gamma}$. The angles $\phi_{\gamma}=0, \tfrac{2 \pi }{3}, \tfrac{4 \pi }{3}$ refer to the $z$, $x$, and $y$ type bonds, respectively. The transformations between the two sets of parameters entering Eq.~\ref{eq:HK} and Eq.~\ref{eq:H} are:
\begin{alignat}{2}
&J_{XY} =J+\tfrac{1}{3}K-\tfrac{1}{3}(\Gamma+2\Gamma') \; ,
&\quad \ \ \ \ \ \ \ \ \ \ \ \ \
&K =A+2B \; ,
\notag \\
&J_Z =J+\tfrac{1}{3}K+\tfrac{2}{3}(\Gamma+2\Gamma')\; , &
&J =\tfrac{1}{3}(2J_{XY}+J_Z-A-2B) \; ,
\notag \\
&A =\tfrac{1}{3}K+\tfrac{2}{3}(\Gamma-\Gamma') \; , &
&\Gamma =\tfrac{2}{3}(A-B)+\tfrac{1}{3}(J_Z-J_{XY})  \; ,
\notag \\
&B =\tfrac{1}{3}K-\tfrac{1}{3}(\Gamma-\Gamma')  \; , &
&\Gamma' =\tfrac{1}{3}(J_Z-J_{XY}+B-A) \; .
\label{eq:Tr}
\end{alignat}

Since the pseudospin wavefunctions (\ref{eq:wf}) are defined in the trigonal
$XYZ$ basis, it is technically simpler to derive $\widetilde{S}=1/2$ Hamiltonian in a form of (\ref{eq:H}), and then convert the results onto a cubic $xyz$ reference frame via Eqs.~\ref{eq:Tr}.

As discussed in the main text, there are three basic exchange channels in $d^7$ systems, which we consider now in detail. General form of the Kugel-Khomskii type spin-orbital Hamiltonians were obtained earlier \cite{Liu18}; for completeness, they will be reproduced below. Here, the major task is to derive the corresponding pseudospin-1/2 Hamiltonians in a realistic case of finite trigonal splitting of $t_{2g}$ orbitals. As the $\widetilde{S}=1/2$ wavefunctions (\ref{eq:wf}) are somewhat complicated, the calculations are tedious but can still be done
analytically.

%=======================================================================
\subsection{1. $t_{2g}$-$t_{2g}$ exchange contributions}
%=======================================================================

%-----------------------------------------------------------------------
\subsubsection{1.1 Intersite $U$ processes}
%-----------------------------------------------------------------------
The spin-orbital Hamiltonian for these exchange processes is given
by equations (A2) and (3) of Ref.~\cite{Liu18}:
\begin{align}
\mathcal{H}^{(z)}_{11} =
&\frac{4t^2}{9}\frac{1}{E_1}(\vc S_i \cdot \vc S_j+S^2)
(a_i^{\dag}b_ia_j^{\dag}b_j+b_i^{\dag}a_ib_j^{\dag}a_j)  \notag \\
+&\frac{4 t^2}{27} \left(\frac{1}{E_3}+\frac{2}{E_2}\right)
(\vc S_i \cdot \vc S_j+S^2)(n_{ia}n_{jb}+n_{ib}n_{ja})  \notag \\
-&\frac{t^2}{6}\left(\frac{1}{E_1}\!-\!\frac{1}{E_2}\right)
(\vc S_i \cdot \vc S_j\!+\!S^2)
\left[(n_{ia}\!-\!n_{jb})^2\!+\!(n_{ib}\!-\!n_{ja})^2\right]\notag \\
-&\frac{4 t^2}{27}\left(\frac{1}{E_2}-\frac{1}{E_3}\right)
(\vc S_i \cdot \vc S_j-S^2)
(a_i^{\dag}b_ib_j^{\dag}a_j+b_i^{\dag}a_ia_j^{\dag}b_j)  \notag \\
+&\frac{t^2}{6}\left(\frac{3}{E_1}+\frac{1}{E_2}-\frac{4}{E_3}\right)
(n_{ia}n_{jb}+n_{ib}n_{ja})  \notag \\
-&\frac{4}{9}\frac{tt'}{U}(\vc S_i \cdot \vc S_j+S^2)
\left[(a_i^{\dag}c_ic_j^{\dag}b_j
+c_i^{\dag}a_ib_j^{\dag}c_j)+(a\leftrightarrow b)\right] \notag \\
+&\frac{4}{9}\frac{t'^2}{U}(\vc S_i \cdot \vc S_j-S^2)\;n_{ic}n_{jc}.
\label{eq:JHA1}
\end{align}
Here $n_a=a^{\dag}a$, etc. denote the orbital occupations, $t$ is the hopping between $a$ and $b$ orbitals via ligand ions, $t'$ is the direct overlap of $c$ orbitals. The Mott-Hubbard excitation energies are $E_1=U-3J_H$, $E_2=U+J_H$, and $E_3=U+4J_H$, where $U$ and $J_H$ are Coulomb repulsion and Hund's coupling on Co$^{2+}$ ions.

Now, we need to express various combinations of the spin and orbital operators above in terms of the pseudospins $\widetilde{S}=1/2$ defined by Eq.~\ref{eq:wf}.
To this end, we have derived a general projection table, presented in subsection 4 below. Using this table, we obtain the pseudospin Hamiltonian in the form of Eq.~\ref{eq:H}, with the following exchange constants:
\begin{align}
J^{XY}_{11} = &\frac{4t^2}{27}\left( \frac{3}{E_1}-\frac{1}{E_2}
+\frac{1}{E_3} \right) \left( 2u_4^2 + 2 u_6 ^2 -\tfrac{13}{2}u_5^2\right)
+ \frac{t^2}{27}\left( \frac{9}{E_1}-\frac{1}{E_2}
+ \frac{4}{E_3} \right) \left( \tfrac{2}{9} u_1^2 - u_4 ^2 -\tfrac{1}{2}u_5^2\right)
\notag \\
-& \frac{2t^2}{9} \left( \frac{1}{E_1}-\frac{1}{E_2}\right) u_1^2
- \frac{4}{9} \frac{tt'}{U}  \left( 4u_6^2 - 2 u_4 ^2 +\tfrac{13}{2}u_5^2\right)
+ \frac{4}{9} \frac{t'^2}{U}  \left( \tfrac{1}{9} u_1^2 + u_4 ^2 +\tfrac{1}{2}u_5^2\right) ,
\\
J^Z_{11} = &\frac{4t^2}{27}\left( \frac{3}{E_1}-\frac{1}{E_2}
+\frac{1}{E_3} \right) \left[ 2u_7^2 + u_3 ^2 -\tfrac{3}{8}(u_2-1)^2\right]
+\frac{t^2}{27}\left( \frac{9}{E_1}-\frac{1}{E_2}
+\frac{4}{E_3} \right) \left( \tfrac{2}{9} u_2^2 - 2u_3 ^2\right)
\notag \\
-& \frac{2t^2}{9} \left( \frac{1}{E_1}-\frac{1}{E_2}\right) u_2^2
- \frac{4}{9} \frac{tt'}{U}  \left[ 4u_7^2 - u_3 ^2 -\tfrac{3}{4}(u_2-1)^2 \right]
+ \frac{4}{9} \frac{t'^2}{U}  \left( \tfrac{1}{9} u_2^2 + 2u_3 ^2 \right) ,
\notag \\
A_{11} = &-\frac{4t^2}{27}\left( \frac{3}{E_1}-\frac{1}{E_2}
+\frac{1}{E_3} \right) \left( 4 u_4 u_6+\tfrac{13}{2}u_5^2\right)
+\frac{t^2}{27}\left( \frac{9}{E_1}-\frac{1}{E_2}
+\frac{4}{E_3} \right) \left(\tfrac{2}{3} u_1 u_4+u_5^2\right)
\notag \\
-& \frac{t^2}{3} \left( \frac{1}{E_1}-\frac{1}{E_2}\right) u_1 u_4
- \frac{4}{9} \frac{tt'}{U}  \left( 4 u_4 u_6 - 13u_5^2 \right)
+ \frac{4}{9} \frac{t'^2}{U}  \left( \tfrac{1}{2} u_5^2 -\tfrac{2}{3} u_1 u_4 \right) ,
\notag \\
B_{11} =
&\frac{4t^2}{27}\left( \frac{3}{E_1}-\frac{1}{E_2}
+\frac{1}{E_3} \right) \left[ u_3(u_6-u_4)-\frac{u_5}{\sqrt{2}}
\left(u_7-\tfrac{9}{4}u_2+\tfrac{9}{4}\right)\right]
+\frac{t^2}{27}\left( \frac{9}{E_1}-\frac{1}{E_2}
+\frac{4}{E_3} \right)  \left[ u_3 \left(\tfrac{1}{3} u_1+2u_4 \right)-
\tfrac{1}{3\sqrt{2}}u_2u_5\right]
\notag \\
-& \frac{t^2}{6} \left( \frac{1}{E_1}-\frac{1}{E_2}\right) \!
\left( u_1 u_3 - \tfrac{1}{\sqrt{2}} u_2 u_5 \right)
+ \frac{4}{9} \frac{tt'}{U}  \left[ u_3(2u_4+u_6)-\frac{u_5}{\sqrt{2}}
\left(u_7-\tfrac{9}{4}u_2+\tfrac{9}{4} \right) \right]
- \frac{4}{9} \frac{t'^2}{U} \! \left[ u_3 \left(\tfrac{1}{3}u_1-u_4 \right) \notag
 - \tfrac{1}{3\sqrt{2}} u_2 u_5 \right].
\end{align}
Coefficients $u_i$ ($i=1,2,\ldots,7$) are given by Eqs.~\ref{eq:uva} below;
they depend on the spatial shape of the pseudospin wavefunctions (\ref{eq:wf}),
and thus decide how the relative values of the pseudospin interactions vary as a function
of trigonal field $\Delta$.

%-----------------------------------------------------------------------
\subsubsection{1.2 Charge-transfer processes}
%-----------------------------------------------------------------------
The spin-orbital Hamiltonian is (Eq.~9 in Ref.~\cite{Liu18}):
\begin{align}
\mathcal{H}^{(z)}_{12} & = \frac{4}{9}\frac{t^2}{\Delta_{pd}+\frac{U_p}{2}}
(\vc S_i \cdot \vc S_j-S^2)(n_{ia}n_{jb}+n_{ib}n_{ja})
-\frac{2}{9}\frac{t^2 \; J^p_H}{(\Delta_{pd}+\frac{U'_p}{2})^2}\;
\vc S_i \cdot \vc S_j(n_{ic}+n_{jc}),
\label{eq:12}
\end{align}
where $\Delta_{pd}$ is charge-transfer gap. $U_p$ and $U'_p=U_p-2J^p_H$ are the intra- and inter-orbital
Coulomb repulsion of the ligand $p$ orbitals, respectively, and $J^p_H$ is
the Hund's coupling.

Using the projection table of subsection 4, we find the exchange constants in the form of Eq.~\ref{eq:H}:
\begin{align}
J^{XY}_{12} & = \frac{4}{9}\frac{t^2}{\Delta_{pd}+\frac{U_p}{2}}
\left( \tfrac{2}{9}u_1^2 - u_4 ^2 -\tfrac{1}{2}u_5^2\right)
-\frac{4}{27}\frac{t^2 \; J^p_H}{(\Delta_{pd}+\frac{U'_p}{2})^2}\; u_1^2 \; ,
\notag \\
J^Z_{12} & = \frac{4}{9}\frac{t^2}{\Delta_{pd}+\frac{U_p}{2}}
\left( \tfrac{2}{9}u_2^2 - 2u_3 ^2\right)
-\frac{4}{27}\frac{t^2 \; J^p_H}{(\Delta_{pd}+\frac{U'_p}{2})^2}\; u_2^2 \; ,
\notag \\
A_{12} & = \frac{4}{9}\frac{t^2}{\Delta_{pd}+\frac{U_p}{2}}
\left( \tfrac{2}{3}u_1 u_4 + u_5 ^2\right)
+\frac{4}{9}\frac{t^2 \; J^p_H}{(\Delta_{pd}+\frac{U'_p}{2})^2}\; u_1u_4 \; ,
\notag \\
B_{12} &  = \frac{4}{9}\frac{t^2}{\Delta_{pd}+\frac{U_p}{2}}
\left[ u_3 \left(\tfrac{1}{3}u_1+2u_4 \right)
 - \tfrac{1}{3\sqrt{2}} u_2 u_5 \right]
+\frac{2}{9}\frac{t^2 \; J^p_H}{(\Delta_{pd}+\frac{U'_p}{2})^2}\;
\left(u_1u_3-\tfrac{1}{\sqrt{2}} u_2 u_5\right).
\end{align}

%-----------------------------------------------------------------------
\subsubsection{1.3 Cyclic exchange processes}
%-----------------------------------------------------------------------
The spin-orbital Hamiltonian is (Eq.~11 in Ref.~\cite{Liu18}):
\begin{equation}
\mathcal{H}^{(z)}_{13} = \frac{4}{9}\frac{t^2}{\Delta_{pd}}(\vc S_i \cdot
\vc S_j+S^2)(a_i^{\dag}b_ia_j^{\dag}b_j+b_i^{\dag}a_ib_j^{\dag}a_j).
\label{eq:A3}
\end{equation}

After projection, we obtain the exchange constants as:
\begin{alignat}{2}
J^{XY}_{13} &= \frac{4}{9}\frac{t^2}{\Delta_{pd}}
\left( 2u_4^2 + 2 u_6 ^2 -\tfrac{13}{2}u_5^2\right),
&\quad \ \ \ \ \ \ \ \ \ \ \ \ \
J^Z_{13} &= \frac{4}{9}\frac{t^2}{\Delta_{pd}}
\left[ 2u_7^2 + u_3 ^2 -\tfrac{3}{8}(u_2-1)^2\right] ,
\notag \\
A_{13} &= -\frac{4}{9}\frac{t^2}{\Delta_{pd}}
 \left( 4 u_4 u_6+\tfrac{13}{2}u_5^2\right),
&
B_{13} &=\frac{4}{9}\frac{t^2}{\Delta_{pd}}
 \left[ u_3(u_6-u_4)-\frac{u_5}{\sqrt{2}}
\left(u_7-\tfrac{9}{4}u_2+\tfrac{9}{4}\right)\right].
\end{alignat}

The total contribution from $t_{2g}$-$t_{2g}$ hopping channel to Eq.~\ref{eq:H}
is given by
\begin{alignat}{2}
J^{XY}_1 &=  J^{XY}_{11}+J^{XY}_{12}+J^{XY}_{13} \;,
&\quad \ \ \ \ \ \ \ \ \ \ \ \ \
J^Z_1 &=  J^Z_{11}+J^Z_{12}+J^Z_{13} \;,
\notag \\
A_1 &= A_{11}+A_{12}+A_{13} \;,
&
B_1 &= B_{11}+B_{12}+B_{13} \;.
\label{eq:tt}
\end{alignat}
The corresponding $K$, $J$, $\Gamma$, and $\Gamma'$ values can be obtained using Eqs.~\ref{eq:Tr}.

%=======================================================================
\subsection{2. $t_{2g}$-$e_g$ exchange contributions}
%=======================================================================
%-----------------------------------------------------------------------
\subsubsection{2.1 Intersite $U$ processes}
%-----------------------------------------------------------------------
The corresponding spin-orbital exchange Hamiltonian is (Eq.~A5 in Ref.~\cite{Liu18}):
\begin{align}
\mathcal{H}^{(z)}_{21} &=
\frac{4\alpha_1}{9}\frac{tt_e}{\widetilde{U}}
(\vc S_i \cdot \vc S_j-S^2)(n_{ic}+n_{jc})
-\frac{tt_e}{6}\frac{\Delta_e}{\Delta_{pd}}
\left(\frac{1}{E_1+D}\!-\!\frac{1}{E_2+D} \right)
\vc S_i \cdot \vc S_j \; (2-n_{ic}-n_{jc}).
\label{eq:JHB1}
\end{align}
Here, $t_e=t_{pd \sigma }^2/\Delta_e$, with $t_{pd \sigma}$ representing hopping between $p$ and $e_g$ orbitals via the charge-transfer gap $\Delta_e=\Delta_{pd}+D$.
Parameter $D$ is the splitting between $t_{2g}$ and $e_g$ levels. The constants $\alpha_1$ and $1/\widetilde{U}$ are:
\begin{align}
\alpha_1 = 1-\frac{D^2}{2\Delta_{pd}\Delta_e}
\left(\frac{\Delta_{pd}+\Delta_e}{U+2J_H}-1\right),
\ \ \ \ \ \ \ \ \ \ \ \ \
\frac{1}{\widetilde{U}} = \frac{1}{6}\left(\frac{2}{E_2+D}\!+\!
\frac{1}{E_3+D}\!+\!\frac{3}{U+2J_H-D}\right).
\end{align}

After projection onto pseudospin-1/2 doublet (\ref{eq:wf}), we get the exchange
constants in the form of Eq.~\ref{eq:H}:
\begin{align}
J^{XY}_{21} &= \left[\frac{8\alpha_1}{27}\frac{tt_e}{\widetilde{U}}
-\frac{2tt_e}{9}\frac{\Delta_e}{\Delta_{pd}}
\left(\frac{1}{E_1+D}\!-\!\frac{1}{E_2+D} \right)  \right] u_1^2 \; ,
\notag \\
J^Z_{21} &= \left[\frac{8\alpha_1}{27}\frac{tt_e}{\widetilde{U}}
-\frac{2tt_e}{9}\frac{\Delta_e}{\Delta_{pd}}
\left(\frac{1}{E_1+D}\!-\!\frac{1}{E_2+D} \right)  \right] u_2^2 \; ,
\notag \\
A_{21} & =  -\left[\frac{8\alpha_1}{9}\frac{tt_e}{\widetilde{U}}
+\frac{tt_e}{3}\frac{\Delta_e}{\Delta_{pd}}
\left(\frac{1}{E_1+D}\!-\!\frac{1}{E_2+D} \right)  \right] u_1 u_4 \; ,
\notag \\
B_{21} &= \left[\frac{4\alpha_1}{9}\frac{tt_e}{\widetilde{U}}
+\frac{tt_e}{6}\frac{\Delta_e}{\Delta_{pd}}
\left(\frac{1}{E_1+D}\!-\!\frac{1}{E_2+D} \right)  \right]
\left( \frac{u_2 u_5}{\sqrt{2}}- u_1 u_3\right).
\end{align}

%-----------------------------------------------------------------------
\subsubsection{2.2 Charge-transfer processes}
%-----------------------------------------------------------------------
The spin-orbital Hamiltonian describing these processes is (Eq.~19 in Ref.~\cite{Liu18}):
\begin{align}
\mathcal{H}^{(z)}_{22}& =
\frac{8\alpha_2}{9}\frac{tt_e}{\Delta_{pd}+\frac{U_p}{2}}(\vc S_i \cdot \vc S_j-S^2)
(n_{ic}+n_{jc})
-\frac{2\alpha_3}{9}\frac{tt_e \; J^p_H}{(\Delta_{pd}+\frac{D+U'_p}{2})^2}\;
\vc S_i \cdot \vc S_j \; (2-n_{ic}-n_{jc}),
\label{eq:b2}
\end{align}
where
\begin{align}
\alpha_2 = 1-\frac{D}{4(\Delta_e+\frac{U_p}{2})}
+\frac{D\;U_p}{8\Delta_{pd}(\Delta_e+\frac{U_p}{2})}-\frac{D}{4\Delta_e},
\ \ \ \ \ \ \ \ \ \ \
\alpha_3 =\frac{(\Delta_{pd}+\Delta_e)^2}{4\Delta_{pd}\Delta_e}\;.
\end{align}

The corresponding pseudospin exchange constants are:
\begin{align}
J^{XY}_{22} &= \left[\frac{16\alpha_2}{27}\frac{tt_e}{\Delta_{pd}+\frac{U_p}{2}}
-\frac{8\alpha_3}{27}\frac{tt_e \; J^p_H}{(\Delta_{pd}+\frac{D+U'_p}{2})^2}  \right] u_1^2 \; ,
\notag \\
J^Z_{22} &= \left[\frac{16\alpha_2}{27}\frac{tt_e}{\Delta_{pd}+\frac{U_p}{2}}
-\frac{8\alpha_3}{27}\frac{tt_e \; J^p_H}{(\Delta_{pd}+\frac{D+U'_p}{2})^2}  \right] u_2^2 \; ,
\notag \\
A_{22} &= -\left[\frac{16\alpha_2}{9}\frac{tt_e}{\Delta_{pd}+\frac{U_p}{2}}
+\frac{4\alpha_3}{9}\frac{tt_e \; J^p_H}{(\Delta_{pd}+\frac{D+U'_p}{2})^2}  \right] u_1 u_4 \; ,
\notag \\
B_{22} &= \left[\frac{8\alpha_2}{9}\frac{tt_e}{\Delta_{pd}+\frac{U_p}{2}}
+\frac{2\alpha_3}{9}\frac{tt_e \; J^p_H}{(\Delta_{pd}+\frac{D+U'_p}{2})^2}  \right]
\left( \frac{u_2 u_5}{\sqrt{2}}- u_1 u_3\right).
\end{align}

%-----------------------------------------------------------------------
\subsubsection{2.3 Cyclic exchange processes}
%-----------------------------------------------------------------------
The corresponding spin-orbital Hamiltonian is (Eq.~22 in Ref.~\cite{Liu18}):
\begin{equation}
\mathcal{H}^{(z)}_{23}=-\frac{2\alpha_4}{9}\frac{tt_e}{\Delta_{pd}}
(\vc S_i \cdot \vc S_j+S^2)(n_{ic}+n_{jc}),
\label{eq:B3}
\end{equation}
with $\alpha_4=1-\frac{1}{2}\frac{D}{\Delta_{pd}+D}$.

After projection onto pseudospin-1/2 doublet, we obtain:
\begin{alignat}{2}
&J^{XY}_{23} = -\frac{4\alpha_4}{27}\frac{tt_e}{\Delta_{pd}} u_1^2 \; ,
&\quad \ \ \ \ \ \ \ \ \ \ \ \ \
&J^Z_{23} = -\frac{4\alpha_4}{27}\frac{tt_e}{\Delta_{pd}} u_2^2 \; ,
\notag \\
&A_{23} =\frac{4\alpha_4}{9}\frac{tt_e}{\Delta_{pd}} u_1 u_4 \; ,
&
&B_{23}= -\frac{2\alpha_4}{9}\frac{tt_e}{\Delta_{pd}} \left( \frac{u_2 u_5}{\sqrt{2}}- u_1 u_3\right).
\end{alignat}

The total contribution from $t_{2g}$-$e_g$ exchange channel to Eq.~\ref{eq:H} is given by
\begin{alignat}{2}
J^{XY}_2 &=J^{XY}_{21}+J^{XY}_{22}+J^{XY}_{23}\; ,
&\quad \ \ \ \ \ \ \ \ \ \ \ \ \
J^Z_2 &=J^Z_{21}+J^Z_{22}+J^Z_{23} \; , \notag \\
A_2 &=A_{21}+A_{22}+A_{23}  \; ,
&\quad \ \ \ \ \ \ \ \ \ \ \ \ \
B_2 &=B_{21}+B_{22}+B_{23} \; .
\label{eq:te}
\end{alignat}

%=======================================================================
\subsection{ 3. $e_g$-$e_g$ exchange contribution}
%=======================================================================
\label{sec:ee}
The corresponding Hamiltonian is very simple (see Eq.~27 in Ref.~\cite{Liu18}):
\begin{equation}
\mathcal{H}^{(z)}_{3}=-\frac{4}{9}
\frac{t^2_e \; J^p_H}{(\Delta_e+\frac{U'_p}{2})^2}\;\vc S_i \cdot \vc S_j.
\label{eq:c1}
\end{equation}
Note that no orbital operators are involved in this interaction and thus it has no bond-dependence. This is because $e_g$ doublet hosts two electrons with parallel spins, leaving no $e_g$-orbital degeneracy. After projecting Eq.~\ref{eq:c1} onto pseudospin subspace, we find
\begin{align}
J^{XY}_3 = -\frac{4}{9}
\frac{t^2_e \; J^p_H}{(\Delta_e+\frac{U'_p}{2})^2} u_1^2 \; ,
\ \ \ \ \ \ \ \ \ \ \
J^Z_3 = -\frac{4}{9}
\frac{t^2_e \; J^p_H}{(\Delta_e+\frac{U'_p}{2})^2} u_2^2 \; ,
\label{eq:ee}
\end{align}
while the bond-dependent terms $A_3=B_3=0$. The latter implies that $e_g$-$e_g$ interaction channel supports the $XXZ$-type model. In the cubic reference frame, Eq.~\ref{eq:HK}, this translates into $K=0$ and $\Gamma=\Gamma'$.

Total values of the exchange constants are obtained by summing up $t_{2g}$-$t_{2g}$, $t_{2g}$-$e_g$, and $e_g$-$e_g$ contributions [Eqs.~\ref{eq:tt}, \ref{eq:te}, and \ref{eq:ee}, respectively], and converted into $K$, $J$, $\Gamma$, and $\Gamma'$ using Eqs.~\ref{eq:Tr}.

%=======================================================================
\subsection{4. Projection table}
%=======================================================================
Calculating the matrix elements of spin-orbital operators within the pseudospin $\widetilde{S}=1/2$ doublet (\ref{eq:wf}), we obtain the correspondence:
\begin{align}
S_+ = u_1\widetilde{S}_+ \;,  \ \ \;
S_- = u_1\widetilde{S}_- \;,  \ \ \;
S_Z = u_2\widetilde{S}_Z \;,  \ \ \;
\end{align}
\begin{align}
S_+ n_a &= \sqrt{2} u_3 e^{i\tfrac{2\pi}{3}} \widetilde{S}_Z
+\frac{u_1}{3} \widetilde{S}_+
-u_4 e^{i\tfrac{4\pi}{3}} \widetilde{S}_- \;,
\notag \\
S_- n_a &= \sqrt{2} u_3 e^{-i\tfrac{2\pi}{3}} \widetilde{S}_Z
+\frac{u_1}{3} \widetilde{S}_-
-u_4 e^{-i\tfrac{4\pi}{3}} \widetilde{S}_+ \;,
\notag \\
S_Z n_a &= \frac{u_2}{3} \widetilde{S}_Z
+\frac{u_5}{2} (\widetilde{S}_X-\sqrt{3} \widetilde{S}_Y) \;,
\end{align}
\begin{align}
S_+ n_b &= \sqrt{2} u_3 e^{-i\tfrac{2\pi}{3}} \widetilde{S}_Z
+\frac{u_1}{3} \widetilde{S}_+
-u_4 e^{-i\tfrac{4\pi}{3}} \widetilde{S}_- \;,
\notag \\
S_- n_b &= \sqrt{2} u_3 e^{i\tfrac{2\pi}{3}} \widetilde{S}_Z
+\frac{u_1}{3} \widetilde{S}_-
-u_4 e^{i\tfrac{4\pi}{3}} \widetilde{S}_+ \;,
\notag \\
S_Z n_b &= \frac{u_2}{3} \widetilde{S}_Z
+\frac{u_5}{2} (\widetilde{S}_X+\sqrt{3} \widetilde{S}_Y) \;,
\end{align}
\begin{align}
S_+ n_c &= \sqrt{2} u_3 \widetilde{S}_Z + \frac{u_1}{3} \widetilde{S}_+
- u_4 \widetilde{S}_- \;, \notag \\
S_- n_c &= \sqrt{2} u_3 \widetilde{S}_Z + \frac{u_1}{3} \widetilde{S}_-
- u_4 \widetilde{S}_+ \;, \notag \\
S_Z n_c &= \frac{u_2}{3} \widetilde{S}_Z -u_5 \widetilde{S}_X \;,
\end{align}
\begin{align}
a^{\dagger}b & = \tfrac{i}{2\sqrt{3}}[(1-u_2)\widetilde{S}_Z
-6 u_5 \widetilde{S}_X] \;, \ \ \ \ \ \ \ \
S_+ a^{\dagger}b =  - \frac{u_3}{\sqrt{2}} \widetilde{S}_Z + u_6 \widetilde{S}_+
- u_4 \widetilde{S}_- \;, \notag \\
S_- a^{\dagger}b & = - \frac{u_3}{\sqrt{2} } \widetilde{S}_Z + u_6 \widetilde{S}_-
- u_4 \widetilde{S}_+ \;, \ \ \ \ \ \ \ \
S_Z a^{\dagger}b = u_7 \widetilde{S}_Z + \frac{u_5}{2} \widetilde{S}_X \;,
\end{align}
\begin{align}
b^{\dagger}c & = \tfrac{i}{2\sqrt{3}}[(1-u_2)\widetilde{S}_Z
-3 u_5 (\sqrt{3}\widetilde{S}_Y-\widetilde{S}_X)] \;, \ \ \ \ \ \ \ \
S_+ b^{\dagger}c =  \frac{u_3}{\sqrt{2} } e^{-i\tfrac{\pi}{3}} \widetilde{S}_Z + u_6 \widetilde{S}_+
- u_4 e^{-i\tfrac{2\pi}{3}} \widetilde{S}_- \;, \notag \\
S_- b^{\dagger}c & = \frac{u_3}{\sqrt{2} } e^{i\tfrac{\pi}{3}} \widetilde{S}_Z + u_6 \widetilde{S}_-
- u_4 e^{i\tfrac{2\pi}{3}} \widetilde{S}_+ \;, \ \ \ \ \ \ \ \
S_Z b^{\dagger}c = u_7 \widetilde{S}_Z
- \frac{u_5}{4} (\widetilde{S}_X-\sqrt{3}\widetilde{S}_Y) \;,
\end{align}
\begin{align}
c^{\dagger}a & = \tfrac{i}{2\sqrt{3}}[(1-u_2)\widetilde{S}_Z
+ 3 u_5 (\sqrt{3}\widetilde{S}_Y+\widetilde{S}_X)] \;, \ \ \ \ \ \ \ \
S_+ c^{\dagger}a  =  \frac{u_3}{\sqrt{2} } e^{i\tfrac{\pi}{3}} \widetilde{S}_Z + u_6 \widetilde{S}_+
- u_4 e^{i\tfrac{2\pi}{3}} \widetilde{S}_- \;, \notag \\
S_- c^{\dagger}a & =  \frac{u_3}{\sqrt{2} } e^{-i\tfrac{\pi}{3}} \widetilde{S}_Z + u_6 \widetilde{S}_-
- u_4 e^{-i\tfrac{2\pi}{3}} \widetilde{S}_+ \;, \ \ \ \ \ \ \ \
S_Z c^{\dagger}a  = u_7 \widetilde{S}_Z
- \frac{u_5}{4} (\widetilde{S}_X+\sqrt{3}\widetilde{S}_Y) \;,
\end{align}

The parameters $u_i$ ($i=1,2,\ldots,7$) are determined by the pseudospin wavefunction (\ref{eq:wf}) parameters $\mathcal{C}_{1,2,3}$ as:
\begin{align}
\label{eq:uva}
u_1 &= 2\sqrt{3}\mathcal{C}_1  \mathcal{C}_3
     +2\mathcal{C}_2^2 \;, \ \ \ \ \ \
u_2 = 1+2(\mathcal{C}_1 ^2-\mathcal{C}_3^2) \;, \ \ \ \ \ \
u_3 = \tfrac{2\sqrt{2}}{3}\mathcal{C}_2 \mathcal{C}_3 -
      \sqrt{\tfrac{2}{3}}\mathcal{C}_1 \mathcal{C}_2  \;, \ \ \ \ \ \
u_4 =\tfrac{2}{3} \mathcal{C}_3^2 \;,
\notag \\
u_5 &=\tfrac{2}{3}\mathcal{C}_2 \mathcal{C}_3 \;,
\ \ \ \ \ \ \ \ \ \ \ \ \ \ \ \ \ \
u_6 =\tfrac{2}{3} \mathcal{C}_2^2 -  \tfrac{1}{\sqrt{3}}
     \mathcal{C}_1  \mathcal{C}_3 \;, \ \ \ \ \ \ \ \
u_7 =\tfrac{1}{3} \mathcal{C}_2^2 +\tfrac{1}{6}
     \mathcal{C}_3^2-\tfrac{1}{2} \mathcal{C}_1^2  \;.
\end{align}
In the cubic limit, where $(\mathcal{C}_1, \mathcal{C}_2, \mathcal{C}_3)=(\tfrac{1}{\sqrt{2}}, \tfrac{-1}{\sqrt{3}}, \tfrac{1}{\sqrt{6}})$, they are
\begin{align}
u_1=u_2=\frac{5}{3} \;, \ \ \; u_3=u_4=\frac{1}{9} \;, \ \ \;
u_5=-\frac{\sqrt{2}}{9} \;, \ \ \; u_6=\frac{1}{18} \;, \ \ \;
u_7=-\frac{1}{9} \;.
\end{align}

%=======================================================================
\subsection{5. Microscopic parameters used in the calculations}
%=======================================================================
Apart from an overall energy scale $t^2/U$, a number of microscopic parameters appeared in the above expressions for exchange constants. Hund's coupling $J_H \sim 0.8$ eV follows from optical data in CoO~\cite{Pra59}; cubic splitting $D$ for 3$d$ ions is of the order of $1.0-1.5$~eV. With the {\it ab initio} estimates of $U\sim 5.0-7.8$~eV~\cite{Ani91,Pic98,Jia10}, this gives $J_H/U \sim 0.1-0.2$ and $D/U \sim 0.13-0.30$. Specifically, we set $J_H/U=0.15$ and $D/U=0.20$. Hund's coupling on oxygen is large, $J_{H}^p\sim 1.2-1.6$ eV~\cite{Foy13}, while $U_p$ is about $\sim 4$~eV, so we use the representative values of $J_{H}^p/U_p=0.3$ and $U_p/U=0.7$. We set a direct hopping $t'=0.2 t$ (i.e. smaller than in 5$d$/4$d$ compounds~\cite{Win16}), but this value is nearly irrelevant here since $t_{2g}$-$t_{2g}$ exchange is of minor importance anyway, see Fig.~2 of the main text. A ratio $t_{pd\sigma}/t_{pd\pi}=2$~\cite{Cha08} is used. Regarding $\Delta/\lambda$ and $U/\Delta_{pd}$ values, we vary them rather broadly, as they most sensitively control the exchange interactions. With the above input parameters, we arrive at $K$, $J$, $\Gamma$, and $\Gamma'$ values presented in the main text. We have verified that while variations of the input parameters result in some changes of the exchange constants, they do not affect the overall picture and conclusions.

%=======================================================================
\section{III. Exact diagonalization: Phase diagrams based on
static correlations and coherent-state analysis}
%=======================================================================

We consider the nearest-neighbor (NN) interaction model (Eq.~2 of the main text or \ref{eq:HK} in the previous section), supplemented by the third-NN Heisenberg exchange $J_3$ that appears as the
major one among the long-range interactions in ab-initio studies~\cite{Win16}.
In this section we show the full evolution of the phase diagram with the
parameter $J_3$ and also demonstrate the robustness of our picture with
respect to variations of the Hund's exchange $J_H$. The data presented here
complements Fig.~1 and Fig.~3(e,f) of the main manuscript.

To determine the magnetic state, we have performed exact diagonalization
using the values of exchange parameters derived in Sec.~II. Utilizing the
Lanczos method, we have obtained exact ground states of the exchange
Hamiltonian for a symmetric, hexagon-shaped cluster containing 24 sites.
Periodic boundary conditions were applied, corresponding to a periodic tiling
of an infinite lattice. Since the small cluster does not allow for spontaneous
symmetry breaking, we inspect its magnetic state by analyzing the static spin
correlations and by employing the method of coherent spin states introduced in
Ref.~\cite{Cha16}.

We focus on real-space correlations that enable us to judge the extent of the
Kitaev spin liquid phase which should be characterized by vanishing
correlations beyond nearest neighbors. By evaluating the static spin
correlations in momentum space, we would be able to detect the magnetically
ordered states that show peaks at the characteristic momenta of the particular
ordering pattern. Here, however, it is favorable to utilize the method of
coherent spin states that provides a better access to the magnetic order
encoded in the complex cluster wavefunction. In essence, it constructs ``classical'' states (coherent spin states) with
spins pointing in prescribed directions and identifies a ``classical'' state
having maximum overlap with the exact cluster ground state. Thanks to its full
flexibility in the individual spin directions, the method can precisely
determine both collinear patterns as well as non-collinear ones. The
``classical'' trial state is a product state of spins pointing in prescribed
directions (in the sense of finding spin up with 100\% probability when
measuring in that particular direction) and as such it excludes quantum
fluctuations. The maximum overlap is therefore a useful indicator of the
amount of quantum fluctuations. For a fluctuation-free state and
non-degenerate cluster ground state, the corresponding probability reaches the
value $1/\text{(number of degenerate patterns)}$. In contrast, Kitaev spin
liquid is highly fluctuating and does not contain a pronounced ``classical''
state which leads to a tiny maximum overlap (see~\cite{Cha16,Rus19} for
details).

\begin{figure}
\begin{center}
\includegraphics[width=15.5cm]{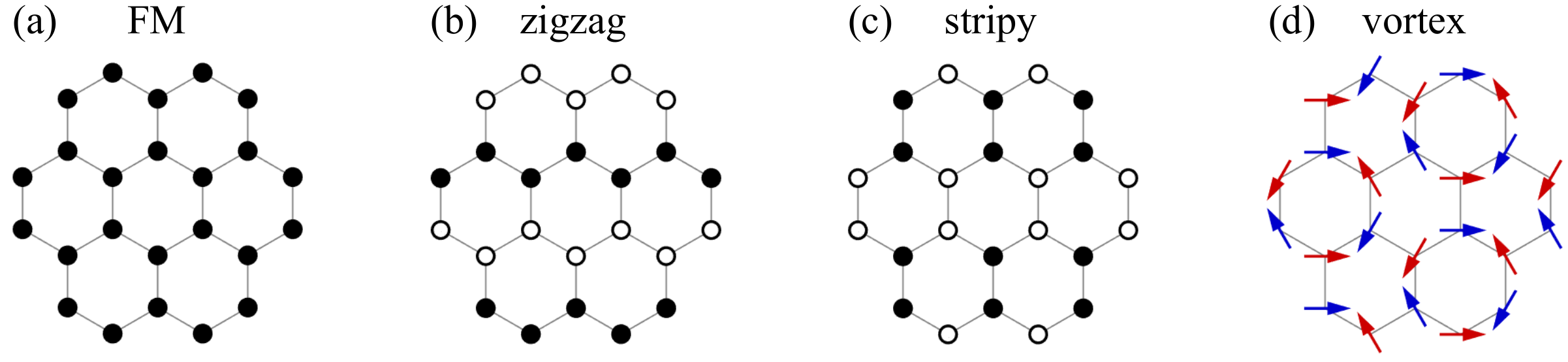}
\caption{Sketch of the magnetic structures for (a) FM, (b) zigzag, (c) stripy, and (d) vortex orders. Open and closed circles represent opposite spin directions.}
\label{fig:s2}
\end{center}
\end{figure}

\begin{figure}
\begin{center}
\includegraphics[width=15.5cm]{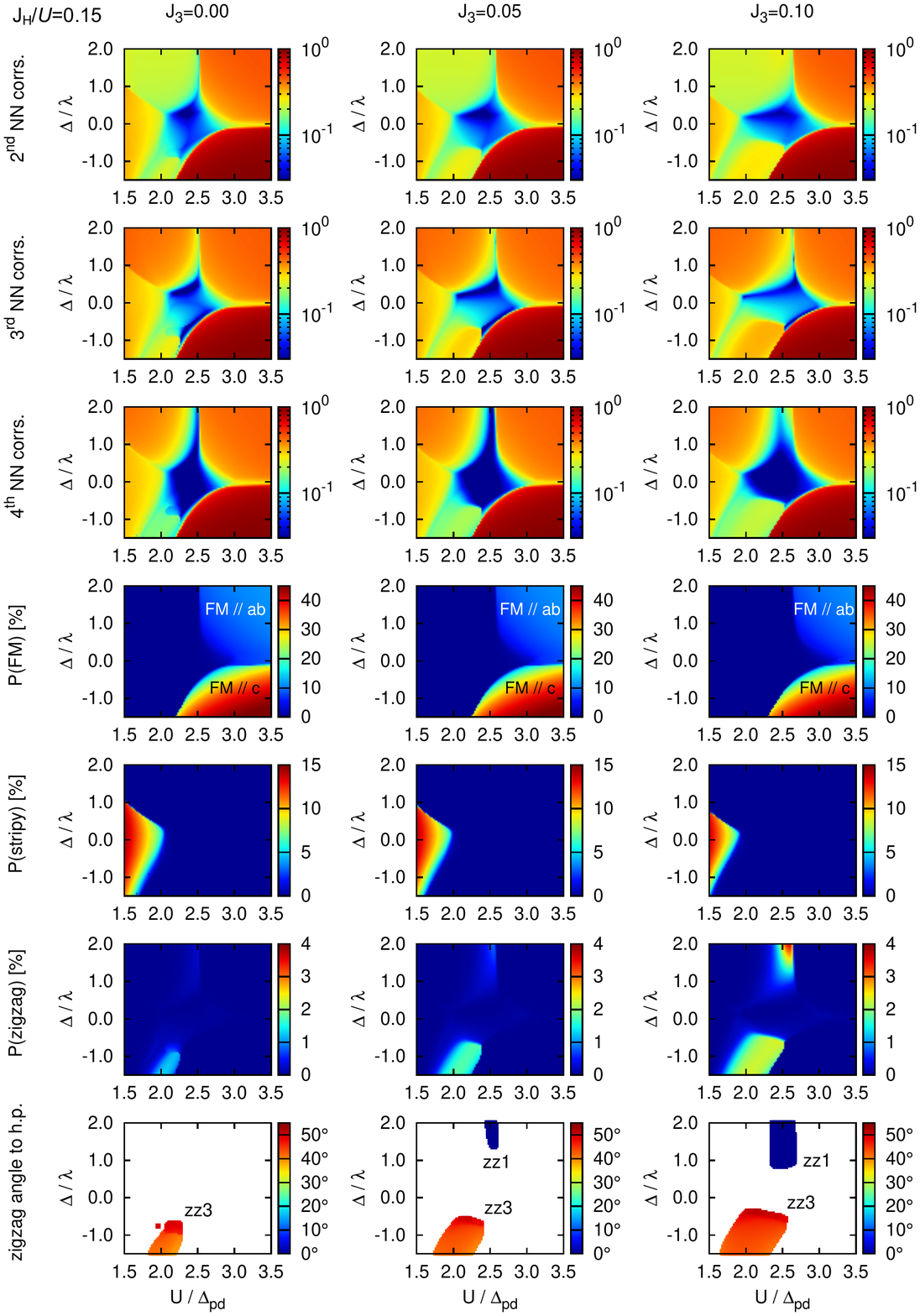}
\caption{The first three rows present second-NN, third-NN and fourth-NN
spin correlations. The color indicates the largest absolute value among
the eigenvalues of the $3\times3$ spin correlation matrix for the respective
bond. It is normalized by the maximum possible value of $\widetilde{S}^2=0.25$.
The next three rows are the probability of FM, stripy, and zigzag classical
states contained in the cluster ground state as determined by the method of
coherent spin states. The last row shows the angle between the honeycomb plane (h.p.) and the magnetic moments for the zigzag order. $J_H/U=0.15$ is fixed and three columns correspond to $J_3=0$, $J_3=0.05$, and $J_3=0.1$ in units of $t^2/U$.}
\label{fig:s3}
\end{center}
\end{figure}

\begin{figure}
\begin{center}
\includegraphics[width=15.5cm]{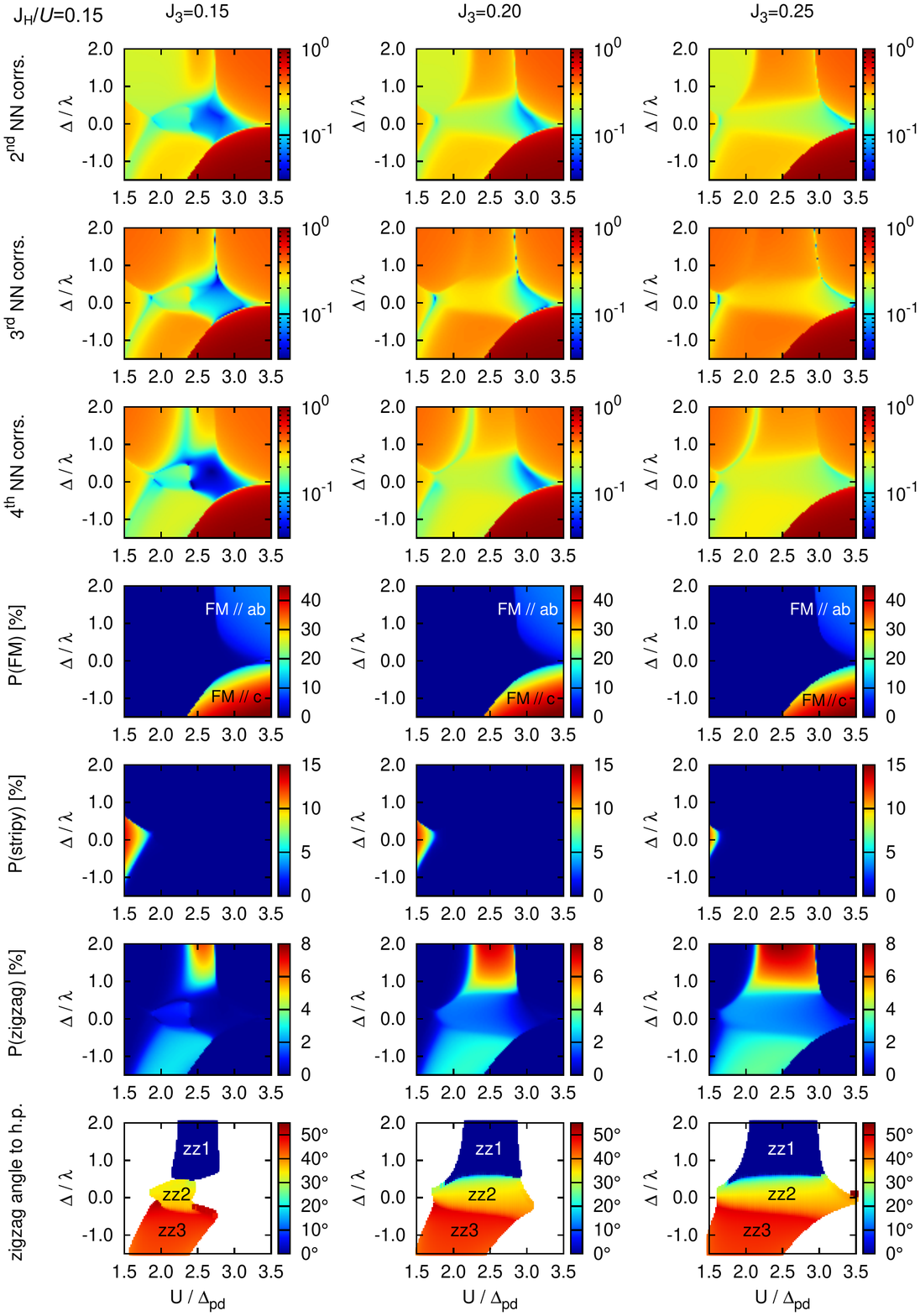}
\caption{The same as in Fig.~\ref{fig:s3} for larger $J_3$ values. The three columns correspond to $J_3=0.15$, $0.20$, and $0.25$ ($t^2/U$).}
\label{fig:s4}
\end{center}
\end{figure}

Figures~\ref{fig:s3} and \ref{fig:s4} show phase diagram data as functions of
$U/\Delta_{pd}$ and $\Delta/\lambda$ for several values of $J_3$.
The static correlations up to fourth NN presented in upper
three rows of panels clearly localize the Kitaev spin liquid phase spreading
in the area with dominant $K$. It is surrounded by several phases with
long-range correlations that are identified by the method of coherent spin
states. For $J_3=0$, these include two types of FM orders with the magnetic
moments lying in the honeycomb plane and perpendicular to it, respectively,
stripy phase, zigzag phase zz3, and finally a vortex phase of the type
depicted in Fig.~\ref{fig:s2}.

%--------------------------------------------------------------------------
%--------------------------------------------------------------------------

The effect of nonzero antiferromagnetic $J_3$ may be estimated by considering
the correlations of third NN in the individual phases. Strongly
supported by $J_3$ is the zigzag phase that is characterized by AF oriented
spins on all third-neighbor bonds. Similarly, a large suppression may be
expected for FM and stripy phases that have FM aligned third NN spins.
The effect on the vortex phase is weak as each spin has one FM aligned third
neighbor and two third neighbors at an angle of $120^\circ$, leading to a
cancellation of $J_3$ in energy on classical level.  Finally, in the Kitaev
spin liquid phase the third neighbors are not correlated at all,
so that small $J_3$ has a moderate negative impact when trying to align
them in AF fashion. The consequences of the above energetics are well visible in
Figs.~\ref{fig:s3} and \ref{fig:s4}. Once including nonzero $J_3$, the Kitaev
spin liquid phase slightly grows first, at the expense of FM and stripy phases.
At the same time, the Kitaev spin liquid phase is also being expelled from
the bottom left corner by the expanding zz3 phase. With increasing $J_3$
between $J_3=0.05$ and $0.15$ in $t^2/U$ units, two new zigzag phases zz1 and
zz2 around Kitaev SL are successively formed. Once $J_3$ reaches $0.25t^2/U$,
the zigzag order quickly takes over, suppressing the Kitaev SL phase completely.

In the large area covered by the zigzag order, various ratios and combinations
of signs of the nearest-neighbor interactions are realized. This is the origin
of three distinct zigzag phases zz1, zz2, and zz3, differing in their moment
directions as seen in bottom panels of Figs.~\ref{fig:s3} and \ref{fig:s4}.
Negative $\Gamma$ and positive $\Gamma'$ found in zz1 phase space
[see Fig.~3(c,d) of the main text] lead to the $ab$-plane moment direction.
The zz3 phase is characterized by opposite signs of $\Gamma$ and $\Gamma'$ interactions which stabilizes the zigzag order as in Na$_2$IrO$_3$~\cite{Chu15,Cha16}. Finally, in the zz2 phase, $\Gamma$ and $\Gamma'$ terms maintain only small
values and moment directions pointing along cubic axes $x$, $y$, $z$ are
selected by order-from-disorder mechanism~\cite{Cha16}.

\begin{figure}
\begin{center}
\includegraphics[width=15.5cm]{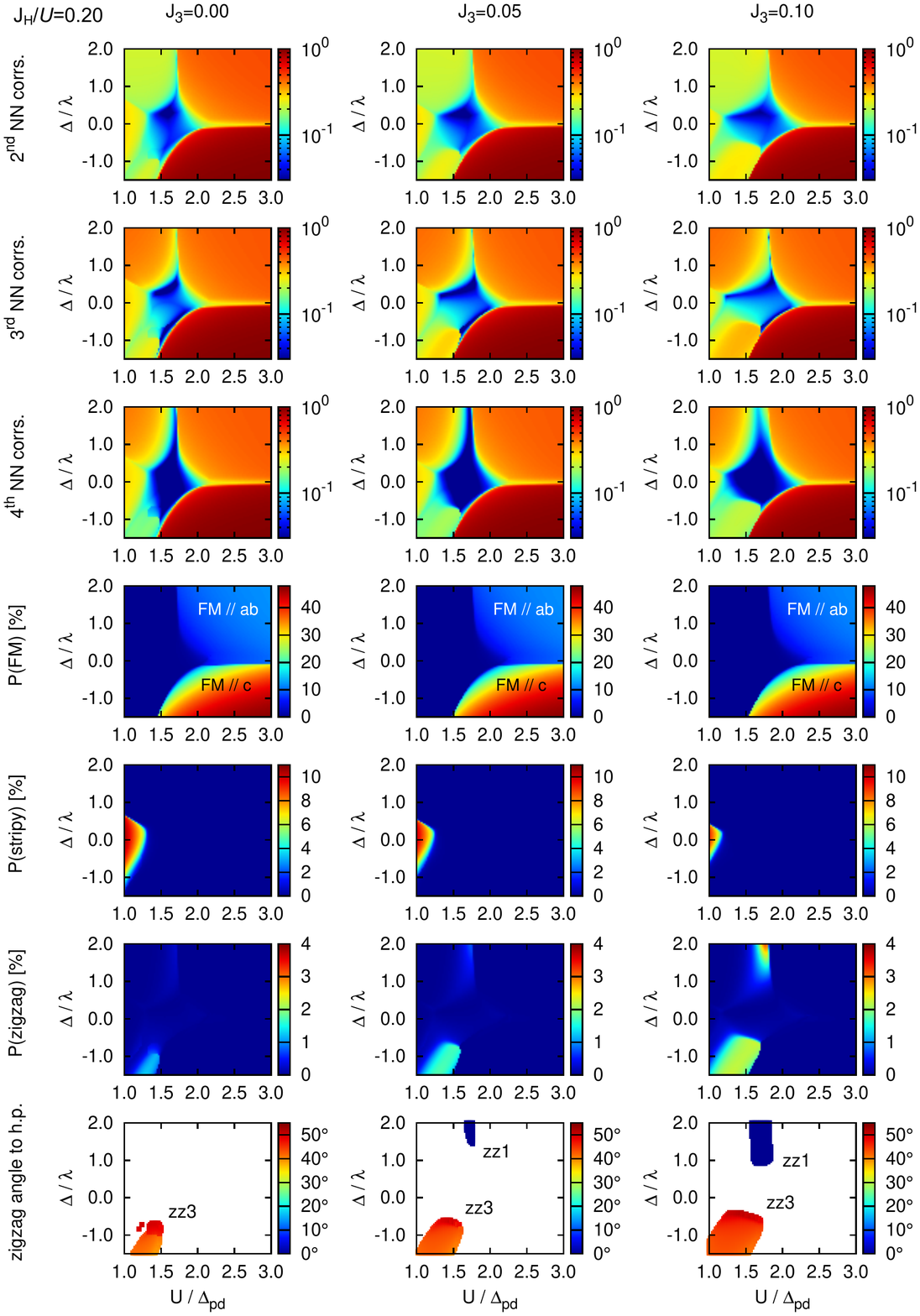}
\caption{The same as in Fig.~\ref{fig:s3} for a larger value $J_H/U=0.2$. The three columns correspond to $J_3=0$, $J_3=0.05$, and $J_3=0.1$ ($t^2/U$).}
\label{fig:s5}
\end{center}
\end{figure}

\begin{figure}
\begin{center}
\includegraphics[width=15.5cm]{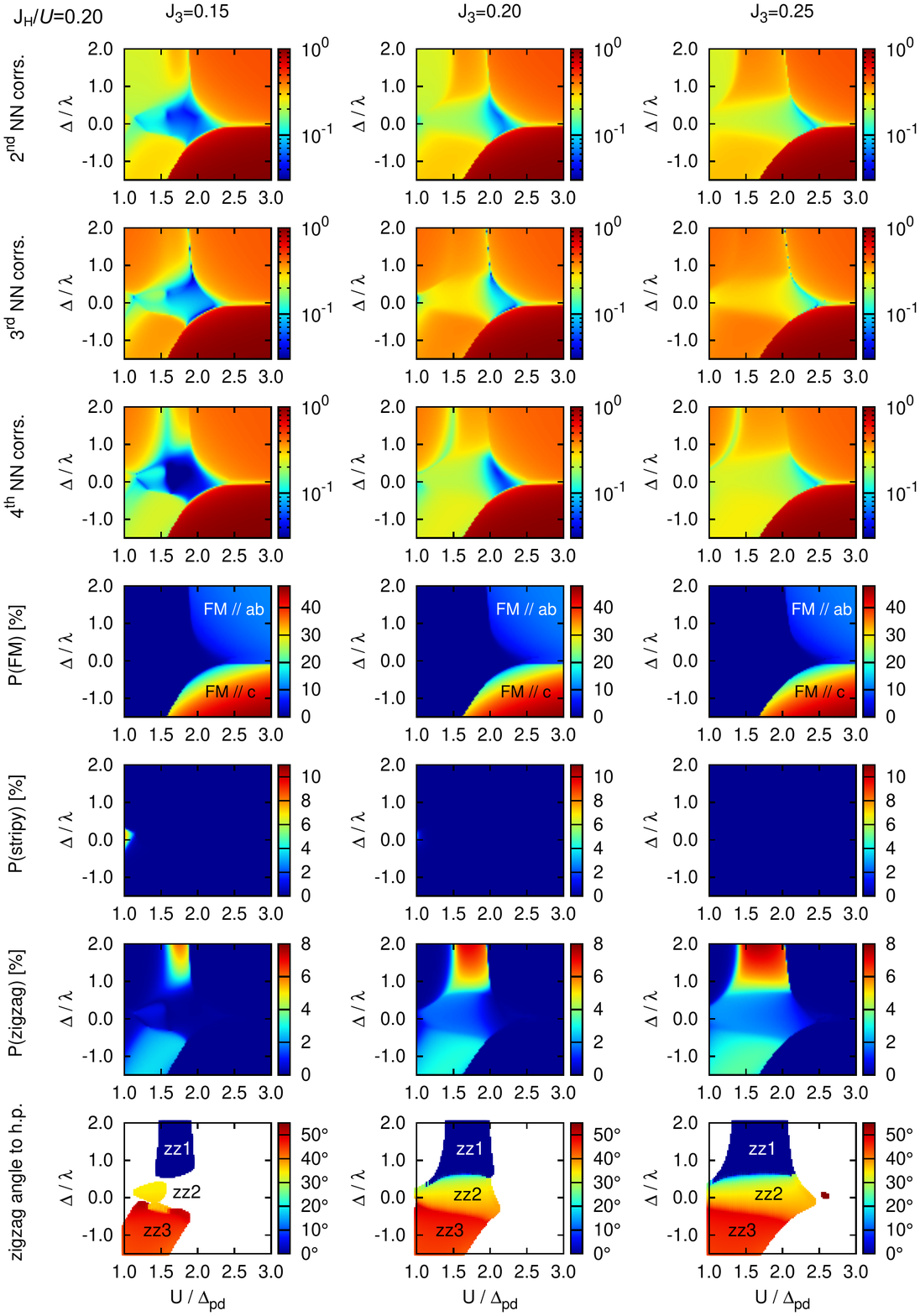}
\caption{The same as in Fig.~\ref{fig:s5} for larger
$J_3$ values. The three columns correspond to $J_3=0.15$, $0.20$, and
$0.25$ ($t^2/U$).}
\label{fig:s6}
\end{center}
\end{figure}

To check the robustness of our picture, we have also performed the exact
diagonalization for a different $J_H$ value. The trends discussed above remain
quite similar as demonstrated in Figs.~\ref{fig:s5} and \ref{fig:s6}
calculated for $J_H/U=0.2$. Roughly speaking, when we increase the $J_H/U$
value, the whole scenario merely shifts to smaller $U/\Delta_{pd}$ region.

%--------------------------------------------------------------------------
%--------------------------------------------------------------------------

%--------------------------------------------------------------------------
%--------------------------------------------------------------------------

%--------------------------------------------------------------------------
%--------------------------------------------------------------------------

%--------------------------------------------------------------------------
%--------------------------------------------------------------------------
\clearpage

%=======================================================================
\section{IV. Trigonal crystal field $\Delta$ in Na$_3$Co$_2$SbO$_6$}
\label{sec:IV}
%=======================================================================

The parameter $\Delta$ determines the effective magnetic moment values $\mu^{\alpha}_{\rm eff}$ ($\alpha=ab$ or $c$), and thus can be obtained from paramagnetic susceptibility $\chi^{\alpha}(T)$. One has to keep in mind that extracting the moments from a standard Curie-Weiss fit $\chi(T)=C/(T-\Theta)+\chi_0$ assumes that the excited levels are high in energy (as compared to $k_BT$) and hence thermally unpopulated. The Curie constant $C$ is then indeed temperature independent, providing the ground state $g$-factors and moments. For Co$^{2+}$ ions, where the excited level at $\sim 30$~meV is thermally activated already at the room temperature, we have to use instead a general expression for a single-ion susceptibility:
\begin{align}
\label{eq:chi}
\chi^{\alpha}_{\rm ion}=\frac{1}{Z(T)}\sum_{n,m}\frac{e^{-\beta E_n}-
e^{-\beta E_m}}{E_m-E_n} (M_{nm}^\alpha)^2 .
\end{align}
Here, $n$ and $m$ run over all the 12 states (6 doublets in Fig.~\ref{fig:s1}), with the wavefunctions and energies calculated in Sec.~I. The partition function $Z(T)= \sum_{n}e^{-\beta E_n}$, and $\beta=1/k_BT$. $M_{nm}^\alpha=\langle n|M_{\alpha}|m \rangle$ is matrix element of the magnetic moment operator $\vc M= (2\vc S-\tfrac{3}{2}\kappa \vc L)$ (in units of Bohr magneton $\mu_B$). We use the covalency reduction factor $\kappa=0.8$ typical for Co$^{2+}$ ion \cite{Abr70}. $\chi^{\alpha}_{\rm ion}$ includes both the Curie and Van-Vleck contributions and depends on two parameters, $\Delta$ and $\lambda$.

\begin{figure}
\begin{center}
\includegraphics[width=15cm]{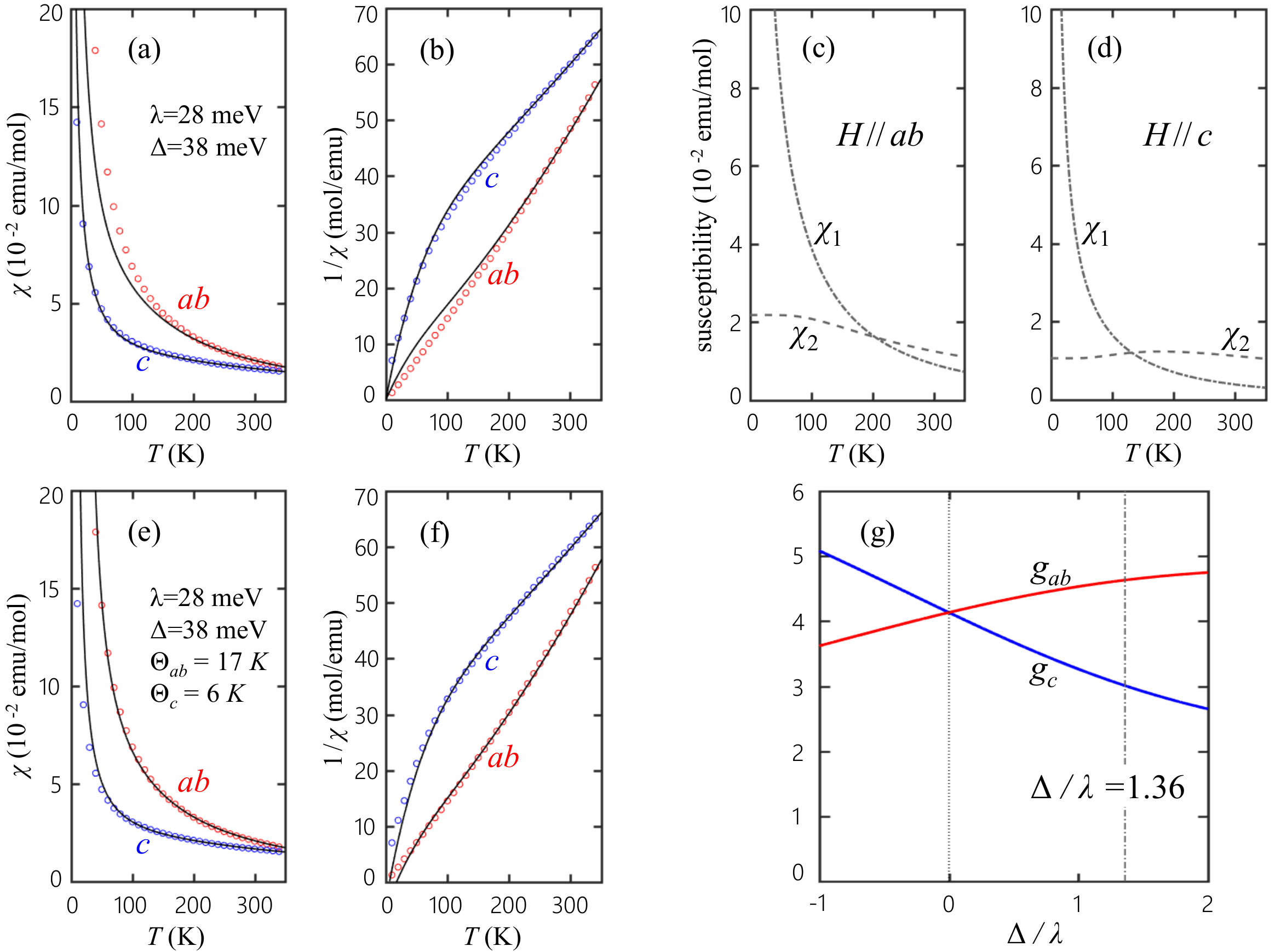}
\caption{(a),(b) Temperature dependence of magnetic susceptibility $\chi$ and its inverse $1/\chi$ in Na$_3$Co$_2$SbO$_6$. Open circles represent the experimental data extracted from Ref.~\cite{Yan19}, and solid lines are the fits using single-ion approximation $\chi^{\alpha}=\chi^{\alpha}_{\rm ion}+\chi^{\alpha}_0$, with $\chi^{ab}_0=-10^{-3}$ emu/mol and $\chi^c_0=1.5\times 10^{-3}$ emu/mol. (c),(d) Decomposition of single-ion susceptibility $\chi^{\alpha}_{\rm ion}$ into pseudospin-1/2 $\chi_1$ and Van-Vleck $\chi_2$ contributions. (e),(f) The fitting results including the pseudospin interactions within a molecular field approximation. Here, $\chi^{ab}_0=-1.5\times 10^{-3}$ emu/mol and $\chi^c_0=1.5\times 10^{-3}$ emu/mol. (g) The g-factors $g_{ab}$ (red) and $g_c$ (blue) as a function of $\Delta/\lambda$. $\Delta/\lambda=1.36$ corresponds to Na$_3$Co$_2$SbO$_6$.}
\label{fig:s7}
\end{center}
\end{figure}

We have fitted the data of Ref.~\cite{Yan19} with $\chi^{\alpha}(T)=\chi^{\alpha}_{\rm ion}+\chi^{\alpha}_0$, and obtained a fair agreement with experiment for both $\chi^{ab}$ and $\chi^c$, using $\Delta=38$~meV and $\lambda=28$~meV, see Fig.~\ref{fig:s7}(a,b). In particular, the characteristic changes in the slopes of both $1/\chi^{ab}$ and $1/\chi^c$ data are well reproduced by the calculations. In fact, this behavior is common for layered cobaltates and deserves some discussion.

It is instructive to divide Eq.~\ref{eq:chi} into two parts, $\chi^{\alpha}_{\rm ion}=\chi^{\alpha}_1+\chi^{\alpha}_2$, where $\chi^{\alpha}_1$ term accounts for the transitions within $\widetilde{S}=1/2$ doublet. Using the wavefunctions (\ref{eq:wf}), we obtain
\begin{align}
\label{eq:chi1}
\chi^{\alpha}_1= p_{1/2} \; \frac{(\mu^{\alpha}_{\rm eff})^2}{3k_B T} \; .
\end{align}
The effective moments $\mu^{\alpha}_{\rm eff}=g_{\alpha} \sqrt{\widetilde{S} (\widetilde{S}+1)}$, with the $\widetilde{S}=1/2$ doublet $g$-factors given by
\begin{align}
g_{ab} &=4\sqrt{3}\mathcal{C}_1\mathcal{C}_3+4\mathcal{C}_2^2
-3\sqrt{2}\kappa\mathcal{C}_2\mathcal{C}_3 \; ,
\notag \\
g_c &=(6+3\kappa)\mathcal{C}_1^2+2\mathcal{C}_2^2
-(2+3\kappa)\mathcal{C}_3^2 \; .
\label{eq:g}
\end{align}
In Eq.~\ref{eq:chi1}, $p_{1/2}=2/Z(T)$ measures the occupation of the ground state. As the excited levels of Co$^{2+}$ are relatively low, the weight $p_{1/2}$ of the Curie term, as well as Van-Vleck contribution $\chi^{\alpha}_2$ of the excited states depend on temperature. The characteristic changes in the slopes of $1/\chi^{ab}$ ($1/\chi^c$) around 200~K (100~K) originate from the interplay between $\chi_1(T)$ and $\chi_2(T)$ which become of a similar order at these temperatures, see Fig.~\ref{fig:s7}(c,d).

The $g$-factors (\ref{eq:g}) are plotted in Fig.~\ref{fig:s7}(g); with $\Delta$ and $\lambda$ values obtained above, we get $g_{ab}\simeq$~4.6 and $g_c\simeq$~3. This gives the in-plane saturated magnetic moment $M_{ab}=g_{ab}\widetilde{S}=2.3\mu_B$ consistent with experiment~\cite{Yan19}.

Apparent deviations at low temperatures are due to short-range correlations between the pseudospins, which can partially be accounted for in a molecular field approximation, i.e. replacing the Curie term $\chi^{\alpha}_1$ by $\chi^{\alpha}_1 \cdot T/(T-\Theta_\alpha)$. The result is shown in Fig.~\ref{fig:s7}(e,f). The paramagnetic Curie temperatures $\Theta_{ab}=17$K and $\Theta_{c}=6$K are rather small and anisotropic. We can evaluate $\Theta$ values using our theoretical exchange constants given in Fig.~4 caption of the main text; the result is:
\begin{align}
\Theta_{ab}=-\tfrac{3}{4}\left [ J+J_3 + \tfrac{1}{3}K-\tfrac{1}{3}(\Gamma+2\Gamma') \right] \simeq 1.4 \;(t^2/U),  \notag \\
\Theta_{c}=-\tfrac{3}{4}\left [ J+J_3 + \tfrac{1}{3}K+\tfrac{2}{3}(\Gamma+2\Gamma') \right] \simeq 0.6 \;(t^2/U).
\label{eq:curie}
\end{align}
Curiously enough, this gives the $\Theta$-anisotropy close to what we get from the susceptibility fits. This comparison also suggests the energy scale of $t^2/U\sim 1$~meV, setting thereby the magnon bandwidth of the order of $10$~meV. The relative smallness of $t^2/U$ is due to large $U$ and more localized nature of $3d$ orbitals.

It is worth to comment on a positive sign of $\Delta>0$ in Na$_3$Co$_2$SbO$_6$. Within a simple model only considering contribution from $O_6$ octahedron, which is slightly compressed along the $c$-axis~\cite{Yan19}, one would find a negative $\Delta<0$ instead. However, this approximation is too crude in layered structures, where the non-cubic Madelung potential of more distant ions has to be considered. In Na$_3$Co$_2$SbO$_6$, we think that $\Delta>0$ is due to a positive contribution of the high-valence $Sb^{5+}$ ions residing within the $ab$-plane. A $c$-axis compression would enhance a negative contribution of the oxygen octahedra, reducing thereby a total value of the trigonal field $\Delta$.

\clearpage

%--------------------------------------------------------------------------
%--------------------------------------------------------------------------

%============================================
\section{V. Dynamical spin susceptibility}
\label{sec:V}
%============================================
\subsection{1. Linear spin wave theory}

The dispersions and intensitites of magnons presented in Fig.~4(a,b) of the main text were determined by standard linear spin wave (LSW) theory. Zigzag pattern with FM $x$ and $y$ bonds was assumed, i.e. the zigzags are running along the $X$ direction in Fig.~S1(a). By applying Holstein-Primakoff transformation, harmonic expansion, and Bogoliubov transformation numerically, we have calculated diagonal components of the spin susceptibility tensor and evaluated its trace that is plotted in Fig.~4(a,b), including artificial lorentzian broadening with FWHM of $0.4$ in units of $t^2/U$.

\subsection{2. Exact diagonalization}

\begin{figure}[h]
\begin{center}
\includegraphics[width=16cm]{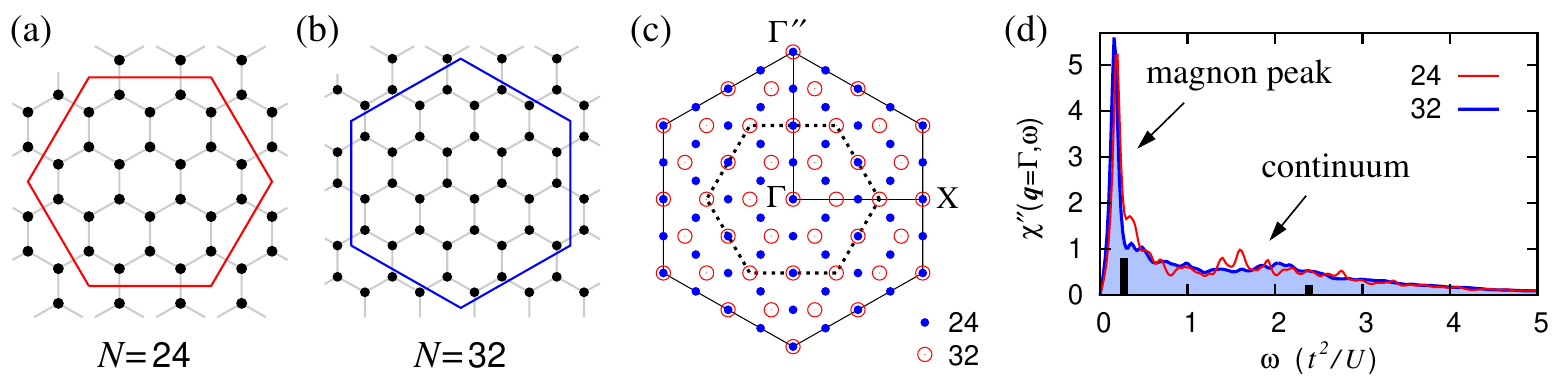}
\caption{
(a) 24-site cluster used in ED to obtain phase diagrams and spin
susceptibility.
(b) 32-site cluster used in ED calculations of the spin susceptibility.
(c) Wavevectors compatible with the periodic tiling of the honeycomb lattice
by 24- and 32-site clusters. Inner dotted hexagon indicates the Brillouin zone
of the honeycomb lattice, outer hexagon corresponds to the Brillouin zone of
the triangular lattice formed when adding sites at hexagon centers to the
honeycomb lattice.
(d) Imaginary part of the trace of the spin susceptibility tensor at $\vc
q=\Gamma=0$ calculated by ED for 24- and 32-site clusters. The values of model
parameters are the same as in Fig.~4 of the main text. The thick black bars
show the positions and relative spectral weights of the magnon peaks obtained
within LSW theory. Note that the ED results for 24- and 32-site clusters are qualitatively similar to each other.
}
\label{fig:EDclust}
\end{center}
\end{figure}

The dynamical spin susceptibility profiles presented in Fig.~4(c) of the main text were determined by exact diagonalization (ED) using the hexagonal clusters with $N=24$ and $N=32$ sites shown in Fig.~\ref{fig:EDclust}(a) and (b), respectively. Utilizing Lanczos algorithm, we have obtained the exact cluster ground state $|\mathrm{GS}\rangle$ and calculated the dynamical spin susceptibility tensor
$\chi_{\alpha\beta}(\vc q,\omega) = i\int \langle\mathrm{GS}| [S_{\vc q}^\alpha(t),S_{-\vc q}^\beta(0)] |\mathrm{GS}\rangle \exp(i\omega t) \theta(t) dt $. Here $S^\alpha_{\vc q}=\sum_{\vc R} S^\alpha_{\vc R}\exp(-i\vc q\vc R)/\sqrt{N}$ is the Fourier component combining spin operators at cluster sites $\vc R$. The accessible wavevectors $\vc q$ that are compatible with periodic tiling of the honeycomb lattice by the clusters are depicted in Fig.~\ref{fig:EDclust}(c). As in the case of the LSW theory, in Fig.~4(c) we have plotted the imaginary part of the trace of the spin susceptibility tensor:
$\chi''(\vc q,\omega)=\mathrm{Im}\sum_\alpha \chi_{\alpha\alpha}(\vc q,\omega)$.
The spectra were broadened by lorentzians with FWHM of $0.1$ in units of $t^2/U$ and the quasielastic peaks at momenta corresponding to the zigzag Bragg points were removed.

%--------------------------------------------------------------------------
%--------------------------------------------------------------------------

Compared to the LSW approximation result, the ED profiles show highly renormalized magnons that only survive at low energies, and broad continua of excitations that emerge as a consequence of the dominant Kitaev interactions. In fact, the most spectral weight is taken by the continuum. This is illustrated in detail for the FM wavevector $\vc q=\Gamma=0$ in Fig.~\ref{fig:EDclust}(d) and can be seen in Fig.~4(c) of the main text for other wavevectors $\vc q$ as well. To properly capture such broad continua, we have used 1000 Lanczos steps in the dynamical susceptibility evaluation.

Finally, we want to notice an important aspect that one has to keep in mind while comparing the above results with the experimental data. Namely, the cluster ground state is fully symmetric and contains all degenerate ordering patterns. In our case these correspond to the three possible zigzag directions that are represented with equal weights for the hexagonal shape clusters. As a result, the dynamical spin susceptibility obtained via ED contains contributions from all these zigzag patterns. In practice, this would correspond to the dynamical spin structure factor measured on the twinned samples with three types of zigzag domains. On the other hand, the intensities calculated using the LSW theory correspond to a single-domain crystal with one particular zigzag pattern.


\begin{thebibliography}{99}
\bibitem{note_axis}
In Fig.~\ref{fig:1}, the $\Delta/\lambda$ axis of Fig.~\ref{fig:3} is replaced by $\Delta$, using $\lambda=28$~meV for Co$^{2+}$ ion.
\bibitem{note_J3}
In fact, the choice of $J_3$ is dictated by the close proximity of zz1 and FM/\!/$ab$ states in Na$_3$Co$_2$SbO$_6$. Classically, they differ by $J-\Gamma+3J_3$; this gives a rough idea of $J_3\sim -J/3$ (as $\Gamma$ is very small).
\end{thebibliography}
\end{document}